\documentclass[11pt]{article}   
\usepackage{amsthm,amsmath}    
\usepackage{amssymb}
\usepackage{graphicx,url}
\usepackage{enumerate}
\usepackage[ruled,section]{algorithm}
\usepackage{algorithmic}

\theoremstyle{plain}
\newtheorem{theorem}{Theorem}[section]
\newtheorem{lemma}[theorem]{Lemma}

\theoremstyle{definition}

\theoremstyle{remark}
\newtheorem{remark}[theorem]{Remark}

\newcommand{\Dep}{\ensuremath{\operatorname{Dep}}}
\newcommand{\MDep}{\ensuremath{\operatorname{MDep}}}
\newcommand{\rank}{\ensuremath{\operatorname{rank}}}
\newcommand{\diag}{\ensuremath{\operatorname{diag}}}

\newcommand{\PP}{\ensuremath{\mathcal{P}}}
\newcommand{\Z}{\ensuremath{\mathbb{Z}}}
\newcommand{\ord}{\ensuremath{\operatorname{ord}}}
\begin{document}
\title{An introspective algorithm for the integer determinant}
\author{Jean-Guillaume Dumas
\and
Anna Urba\'nska
}
\date{}

\maketitle
\begin{center}
Laboratoire Jean Kuntzmann, UMR CNRS 5224\\
Universit\'e Joseph Fourier, Grenoble I\\
BP 53X, 38041 Grenoble, FRANCE.\\
\{Jean-Guillaume.Dumas;Anna.Urbanska\}@imag.fr\\
\url{ljk.imag.fr/membres/{Jean-Guillaume.Dumas;Anna.Urbanska}}\\
\end{center}

\begin{abstract}
We present an algorithm for computing the determinant of an integer matrix
$A$. The algorithm is
{\em introspective} in the sense that it uses several distinct
algorithms that run in a concurrent manner. During the course of
the algorithm partial results coming from distinct methods can
be combined. Then, depending on
the current running time of each method, the algorithm can emphasize a
particular variant.
With the use of very fast modular routines for linear algebra, our
implementation is an order of magnitude faster than other existing
implementations. Moreover, we prove that
the expected complexity of our algorithm is only
%
$O\big(n^3 \log^{2.5}\left(n\|A\|\right) \big) $
{\em bit operations} in the case of random dense matrices, 
where $n$ is the dimension and $\|A\|$
is the largest entry in the absolute value of the matrix. 
\end{abstract}
\section{Introduction}
One has many alternatives to compute the determinant of an integer
matrix. Over a field, the computation of the determinant is tied to
that of matrix multiplication via block recursive matrix
factorizations \cite{Ibarra1982}. On the one hand,
over the integers, a na\"ive
approach would induce a coefficient growth that would render the
algorithm not even polynomial.
On the other hand,
over finite fields, one can nowadays reach the speed of numerical
routines \cite{Dumas2004}.

Therefore, the classical approach over the integers is to
reduce the computation modulo some primes of constant size and to
recover the integer
determinant from the modular computations. For this, at least two
variants are possible: Chinese remaindering and $p$-adic lifting.

The first variant requires either a good {\em a priori} bound on the size of the
determinant or an early termination probabilistic argument
\cite[\S 4.2]{Dumas2005}. It thus achieves an {\em output
 dependant} bit complexity
of $O\big( \log\left(|\det\left(A\right)|\right)\left(n^{\omega}+n^2\log\left(\|A\|\right)\right)\big)$ where $\omega$ is the
exponent of matrix multiplication \footnote{the value of $\omega$ is $3$ for the classical algorithm,
and $2.375477$ for the Coppersmith-Winograd method, see \cite{Coppersmith1987}}. Of course, with the
coefficient growth, the determinant size can be as large as
$O\left(n \log(n\|A\|) \right)$ (Hadamard's bound) thus giving a large worst case
complexity. The algorithm is Monte Carlo type, its deterministic
(always correct) version
exists and has the complexity of
$O\left(\left(n^{\omega+1}+n^3\log(\|A\|)\right)\log(n\|A\|)\right)$ bit operations.

The second variant uses system solving and $p$-adic lifting
\cite{Dixon1982} to get a potentially large factor of the determinant with a
$O\left(n^3 \log^2(n\|A\|) \right)$ bit complexity.
Indeed, every integer matrix is unimodularly equivalent to a diagonal
matrix $S$ equal to $\diag\left(s_1, \dots, s_n\right)$, where $s_i$
divides $s_{i+1}$. This means that there exist integer matrices
$U, V$ with $\det{U},\det{V}=\pm 1$, such that $A=USV$.
The $s_i$ are called the invariant factors of A. In the
presence of several matrices we will also use the notation $s_i(A)$. Solving a
linear system with a random right hand side reveals $s_n$ as the
common denominator of the solution vector entries with high
probability, see \cite{Pan1988,Abbott1999}. 

The idea of \cite{Abbott1999} is thus to combine both approaches,
i.e. to approximate the determinant by system solving
and recover only the remaining part ($\det(A)/s_n$) via Chinese
remaindering. The Monte Carlo version of Chinese remaindering leads to
an algorithm with the {\bf expected} output-dependant bit complexity
of $O\big(n^{\omega} \log\left(|\frac{\det(A)}{s_n}|\right)+n^3\log^2(n\|A\|)\big)$. We use the
notion of the {\bf expected} complexity to emphasize
that it requires
$\mathbf{E}\left(log\left(\frac{s_n}{\tilde{s}_n}\right)\right)$ to be $O(1)$, where
$\tilde{s}_n$ is the computed factor of $s_n$ and $\mathbf{E}$
denotes the expected value computed over all algorithm instances for a
given matrix $A$.

Then G. Villard remarked that at most $O\left( \sqrt{\log\left(|\det\left(A\right)|\right)} \right)$ invariant factors
can be distinct and that in some propitious cases
we can expect that only the last $O\left( \log(n) \right)$ of those
are nontrivial \cite{Eberly2000}. 
This remark, together with a preconditioned $p$-adic solving to compute the $i$-th invariant
factor lead to a $O\left(
n^{2+\frac{\omega}{2}}\log^{1.5}(n\|A\|)\log^{0.5}(n) \right)$ worst case
Monte Carlo algorithm. Without fast matrix multiplication, the
complexity of the algorithm becomes $O\left(n^{3.5}\log^{2.5}(n\|A\|)\log^2(n)\right)$.
The expected number of invariant factors for a set of
matrices with entries chosen randomly and uniformly from the set of consecutive integers 
$\{0,1\dots\dots \lambda-1\}$ can be proven to be $O\left(\log(n)\right)$. Thus,
we can say that the {\bf expected} complexity of the algorithm is
$O\left( n^3 \log^2(n\|A\|) \log(n)\log_\lambda(n)\right)$. Here, the
term {\bf expected} is used in a slightly different context than in
algorithm \cite{Abbott1999} and describes the complexity in the case
where the matrix has a propitious property i.e., the small number of
invariant factors.

In this paper we will prefer to use the notion of the expected rather
than average complexity. Formally, to compute the average complexity
we have to average the running time of the algorithm over all input
and argorithm instances. The common approach is thus to compute the
expected outputs of the subroutines and use them in the complexity
analysis. This allows us to deal easily with complex algorithms with
many calls to subroutines which depend on randomization. The two approaches are equivalent when
the dependency on the expected value is linear, which is often the
case. However, we can imagine more complex cases of adaptive algorithms
where the relation between average and expected complexity is not obvious. Nevertheless, we believe that the evaluation of the
expected complexity gives a meaningful description of the algorithm. We
emphasize the fact, that the propitious input for which the analysis
is valid can often be quickly detected at runtime.

Note that the actual best worst case complexity algorithm for dense matrices is
$O^{\sim}\left( n^{2.7}\log(\|A\|) \right)$, which is
$O^{\sim}\left( n^{3.2}\log(\|A\|) \right)$ without fast matrix multiplication, by
\cite{Kaltofen2005}. We use the notion
$O^{\sim}(N^{\alpha}\log(\|A\|))$, which is
equivalent to $O(N^{\alpha}\log^\beta(N)\log(\|A\|))$ with some
$\beta\geq 0$. Unfortunately, these last two worst case
complexity algorithms, though asymptotically better than \cite{Eberly2000},
are not the fastest for the generic case or for the actually attainable matrix
sizes. The best expected complexity algorithm is the Las Vegas algorithm
of Storjohann \cite{Storjohann2004} which uses an expected number of
$O\left(n^{\omega}\log(n\|A\|)\log^2(n)\right)$ bit operations. In section
\ref{sec:experiments} we compare the performance of this algorithm
(for both certified and not certified variants)
to ours, based on the experimental results of \cite{Storjohann2005}.

In this paper, we propose a new way to extend the idea of
\cite{Saunders2004,Wan2005} to get the last consecutive invariant
factors with high probability in section \ref{sec:extended}.
Then we combine this with the scheme of \cite{Abbott1999}.

This combination is made in an adaptive way. This means that the algorithm will choose the adequate variant at
run-time, depending on discovered properties of its input. More precisely, in section
\ref{sec:algorithm}, we propose an algorithm which uses timings of its
first part to choose the best termination. This particular kind of
adaptation was introduced in \cite{1997:spe:musser} as
introspective; here we use the more specific definition of
\cite{jgd:2006:AHA}.

In section \ref{sec:complexity} we prove that the expected complexity
of our algorithm is \[O\left( n^3 \log^2(n\|A\|)\sqrt{\log(n)}\right)\]
 bit operations in the case of dense matrices, gaining a
$\log^{1.5}(n)$ factor 
compared to \cite{Eberly2000}.

Moreover, we are able to detect the worst cases during the course of
the algorithm and switch to the asymptotically
fastest method.
In general this last switch is not required and we show in section
\ref{sec:experiments} that when used with the very fast modular
routines of \cite{Dumas2002issac,Dumas2004} and the LinBox library
\cite{Dumas2002Linbox}, our algorithm can be an order of magnitude faster than other existing implementations.

A preliminary version of this paper was
presented in the Transgressive Computing 2006
conference \cite{jgd:2006:det}. Here we give better asymptotic
results for the dense case, adapt our algorithm to the sparse case and
give more experimental evidences.
%
\section{Base Algorithms and Procedures}
In this section we present the procedures in more detail and
describe their probabilistic behavior. We start by a brief
description of the properties of the Chinese Remaindering loop (CRA) with
early termination (ET) (see \cite{Dumas2001}), then proceed with
the {\em LargestInvariantFactor} algorithm to compute $s_n$ (see
\cite{Abbott1999, Eberly2000,Saunders2004}).
We end the section with a summary of ideas of Abbott {\em et al.}
\cite{Abbott1999}, Eberly {\em et al.} and Saunders {\em et al.} \cite{Saunders2004}.
%
%
\subsection{Output dependant Chinese Remaindering Loop (CRA)}\label{sec:cra}
CRA is a procedure based on the Chinese remainder
theorem. Determinants are computed modulo several primes $p_i$.
Then the determinant is reconstructed modulo $p_0 \cdots p_{t}$ in the
symmetric range via the Chinese reconstruction. The integer value of the determinant is thus
computed as soon as the product of $p_i$ exceeds $2|\det\left(A\right)|$.
We know that the product is sufficiently big if it exceeds some upper bound
on this value or, probabilistically, if the reconstructed
value remains identical for several successive additions of modular
determinants. The principle of this early termination (ET) is thus to
stop the reconstruction
before reaching the upper bound, as soon as the determinant remains
the same for several steps \cite{Dumas2001}.

Algorithm \ref{alg:cra}
is an outline of a procedure to compute the determinant
using CRA loops with early termination, correctly with probability
$1-\epsilon$. We start with a lemma.
\begin{lemma}\label{lem:cra}
Let $H$ be an upper bound
for the determinant (e.g. $H$ can be the Hadamard's bound:
$|det\left(A\right)| \leq \left(\sqrt{n}\|A\|\right)^n$).
Suppose that distinct primes $p_i$ greater than $l > 0$ are randomly sampled
from a set $P$ with $|P|\geq 2\lceil  \log_{l}\left(H\right)\rceil$. Let $r_t$ be the
value of the determinant modulo $p_0\cdots p_{t}$ computed in the
symmetric range. We have:
\begin{enumerate}[(i)]
\item\label{cra:certification} $r_t=\det\left(A\right)$, if
$t\geq N = \begin{cases}\lceil \log_l\left(|\det\left(A\right)|\right)\rceil & \text{if}~\det\left(A\right)\neq  0\\0&\text{if}~\det\left(A\right)=0\end{cases}$;
\item\label{cra:no_primes} if $r_t\neq \det\left(A\right)$, then there are at most
$R =\lceil \log_l \left(\frac{|\det\left(A\right)-r_t|}{p_0\cdots
p_{t}}\right)\rceil 
$ 
primes $p_{t+1}$ such that
$r_t = \det\left(A\right)$ mod $p_0\cdots p_t p_{t+1}$;
\item\label{cra:on-the-fly}  if $r_t=r_{t+1}=\cdots=r_{t+k}$
and 
$\frac{R'\left(R'-1\right)\dots\left(R'-k+1\right)}{\left(|P|-t-1\right)\dots\left(|P|-t-k\right)}
< \epsilon$, where $R'=\lceil\log_{l}\frac{H+|r_{t}|}{p_{0}p_{1}\dots
p_{t}}\rceil $, 
then 
$\mathcal{P}\left(r_t\neq\det\left(A\right)\right)<\epsilon$.
\item\label{cra:probability} if $r_t=r_{t+1}=\cdots=r_{t+k}$
and
$k \geq \lceil \frac{ \log\left( 1/\epsilon \right) }{ \log\left(P'\right) -
  \log\left(\log_{l}\left(H\right)\right)  } \rceil$, where $P'= |P|-\lceil\log_l\left(H\right)\rceil$,
then $\mathcal{P}\left(r_t\neq\det\left(A\right)\right)<\epsilon$.
\end{enumerate}
\end{lemma}
\begin{proof}
For (\ref{cra:certification}), notice that
$-\lfloor \frac{p_0\cdots p_t}{2}\rfloor\leq r_t < \lceil \frac{p_0\cdots p_t}{2}\rceil$.
Then $r_t=\det\left(A\right)$ as soon as
$p_0\cdots p_t \geq 2|\det\left(A\right)|$.
With $l$ being the lower bound for $p_i$ this reduces to
$t\geq \lceil\log_l{|\det\left(A\right)|}\rceil$ when
$\det\left(A\right)\neq 0$.
\\
For (\ref{cra:no_primes}), we observe that
$\det\left(A\right)=r_t+K p_0\dots p_t$
and it suffices to estimate the number of primes greater than $l$
dividing $K$.
\\
For (\ref{cra:on-the-fly}) we notice that
$k$ primes dividing $K$ are to be chosen with the probability 
$\frac{ {R \choose k} }{ {{|P|-(t+1)} \choose k} }$. Applying the bound $R'$
for $R$ leads to the result.
\\
For (\ref{cra:probability}) we notice that the latter is bounded by
$\left(\frac{R'}{P'}\right)^{k}$ since
$R' \leq \lceil\log_l\left(\frac{2H}{2} \right)\rceil \leq |P'|$. Solving for $k$
the inequality $\left(\frac{R'}{P'}\right)^{k}< \epsilon$ gives the result.
\end{proof}
The two last points of the theorem give the stopping condition for early
termination. The condition (\ref{cra:on-the-fly}) can be computed
on-the-fly (as in Algorithm \ref{alg:cra}). As a default value and
for simplicity (\ref{cra:probability}) can also be used.

\begin{algorithm}\caption{Early Terminated CRA}\label{alg:cra}
\begin{algorithmic}[1]
\REQUIRE $n\times n$ integer matrix $A$.
\REQUIRE $0 < \epsilon < 1$.
\REQUIRE $H$ - Hadamard's bound ($H=\left(\sqrt{n} \|A\|\right)^n$) 
\REQUIRE $l> 0$, a set P of random primes greater than $l$, $|P|\geq 2\lceil\log_l\left(H\right)\rceil$.
\ENSURE The integer determinant of $A$, correct  with probability at
least $1-\epsilon$.
\bigskip
\STATE 
	$i=0$; 
\REPEAT
\STATE Choose uniformly and randomly a prime $p_{i}$ from the set $P$;
\STATE $P = P\backslash \{p_{i}\}$
\STATE \label{cra:ii} Compute $\det\left(A\right)$ mod $p_i$;
\STATE Reconstruct $r_{i}$, the determinant modulo $ p_0\cdots p_{i}$; \hfill // by Chinese
remaindering
\STATE $k = \max\{t: r_{i-t}=\dots=r_i\}$;
$R'=\lceil\log_{l}\frac{H+|r_{i}|}{p_{0}p_{1}\dots p_{i-k}}\rceil $;
\STATE Increment i;
\UNTIL{$\frac{R'\left(R'-1\right)\dots\left(R'-k+1\right)}{\left(|P|-i+k-1\right)\dots\left(|P|-i\right)}
< \epsilon$ or $\prod p_i \geq 2H$}
\end{algorithmic}
\end{algorithm}

To compute the modular determinant in algorithm \ref{cra:ii} we use
the LU factorization modulo $p_i$. Its complexity is  $O\left(n^\omega + n^2\log(\|A\|)\right)$.

Early termination is particularly useful in the case when the computed determinant
is much smaller than the {\em a priori} bound.
The running time of
this procedure is output dependant.
\subsection{Largest Invariant Factor}\label{sec:lif}
A method to compute $s_n$ for integer matrices
was first stated by V. Pan \cite{Pan1988}
and later in the form of the {\em LargestInvariantFactor} procedure (LIF) in
\cite{Abbott1999,Eberly2000,Dumas2001,Saunders2004}.
The idea is to obtain a~divisor of
$s_n$  by
computing a rational solution of the linear systems $Ax=b$.
If $b$ is chosen uniformly and randomly from a sufficiently large set
of contiguous integers,
then the computed divisor can be as close
as possible to $s_n$ with high probability.
Indeed, with $A=USV$, we can equivalently solve $SVx=U^{-1}b$ for
$y=Vx$, and then solve for $x$. As $U$ and $V$ are unimodular,
the least common multiple of the denominators of $x$ and $y$, $d(x)$ and
$d(y)$ satisfies $d(x)= d(y) \arrowvert s_n$.

Thus, solving $Ax=b$ enables us to get $s_n$ with high
probability. The cost of solving using Dixon $p$-adic lifting
\cite{Dixon1982} is $O\left( n^3 \log^2(n\|A\|)+n\log^2(\|b\|)\right)$ as stated by
\cite{Mulders1999}. 

The algorithm takes as input parameters $\beta$ and $r$ which
are used to control the probability of correctness;
$r$ is the number of successive solvings and $\beta$ is the size of the
set from which the values of a random vector $b$ are chosen, i.e. a
bound for $\|b\|$. With
each system solving, the output $\tilde{s}_n$ of the algorithm is
updated as the $\operatorname{lcm}$ of the current solution denominator $d(x)$ and the result
obtained so far.

The following theorem characterizes the probabilistic behavior of the LIF
procedure.
\begin{theorem}\label{thm:lif}
Let $A$ be a $n\times n$ matrix, $H$ its Hadamard's bound, $r$ and
$\beta$ be defined as above. Then the output
$\tilde{s}_n$
of Algorithm {\em LargestInvariantFactor} of \cite{Abbott1999}
is characterized by the following properties.
\begin{enumerate}[(i)]
\item\label{i} If $r=1$, $p$ is a prime, $ l\geq 1$, then $
\mathcal{P}\left(p^{l}\arrowvert\frac{s_n}{\tilde{s}_n}\right)\leq\frac{1}{\beta}\lceil\frac{\beta}{p^l}\rceil;
$
\item\label{ii} if $r=2$, $\beta=\lceil\log\left(H\right)\rceil$
then $
\mathbf{E}\left(\log\left(\frac{s_n}{\tilde{s}_n}\right)\right) = O\left(1\right);
$
\item\label{iii} if $r=2$,
$\beta=6+ \lceil 2 \log \left(H\right)\rceil$
then $s_n=\tilde{s}_n$
with probability at least 1/3;
\item\label{iv} if $r=\lceil 2 \log \left(\log \left(H\right)\right) \rceil $, $\beta \geq 2$ then $
\mathbf{E}\left(\log\left(\frac{s_n}{\tilde{s}_n}\right)\right) = O\left(1\right);
$
\item\label{v} if $r=\lceil\log\left(\log \left(H\right)\right) + \log\left(\frac{1}{\epsilon}\right)\rceil $, $2\mid \beta$ and $\beta \geq 2$
then $s_n=\tilde{s}_n$ with probability at least $1-\epsilon$;
\end{enumerate}
\end{theorem}
\begin{proof}
The proofs of ($\ref{i}$) 
and (\ref{iv}) are in
\cite{Abbott1999}[Thm. 2, Lem. 2]. The proof of (\ref{iii}) is in
\cite{Eberly2000}[Thm. 2.1]. 
To prove (\ref{ii}) we adapt the proof of (\ref{iii}).
The expected value of the under-approximation of $s_n$ is bounded
by the formula
$$
\sum_{p\mid s_n}\sum_{k=1}^{\lfloor\log_p\left(s_n\right)\rfloor} \log\left(p\right)\left(\frac{1}{\beta}\lceil\frac{\beta}{p^k}\rceil\right)^2,
$$
where the sum is taken over all primes dividing $s_n$.
As  $\frac{1}{\beta}\lceil\frac{\beta}{p^k}\rceil$ is bounded by
$\frac{1}{\beta}+\frac{1}{p^k}$ this can be further expressed as
\begin{align*}
&\sum_{p~prime}\sum_{k=1}^{\infty}\log\left(p\right)\frac{1}{p^{2k}}+\frac{2}{\beta}\sum_{p\mid
s_n}\sum_{k=1}^{\infty}\log\left(p\right)\frac{1}{p^k} +
\frac{1}{\beta^2}\sum_{p\mid s_n}\sum_{k=1}^{\lfloor \log_p\left(s_n\right)\rfloor}\log\left(p\right)\leq\\
&\sum_{p~prime}\log\left(p\right)\frac{1}{p^2-1} + \frac{2}{\beta}\sum_{p\mid
s_n}\log\left(p\right)\frac{1}{p-1}+\frac{1}{\beta^2}\sum_{p\mid s_n}\log\left(p\right)\log_p\left(s_n\right)
\\&1.78 + \frac{2\log\left(s_n\right)}{\beta}+\frac{\log^2\left(s_n\right)}{\beta^2}\leq
5 \in O\left(1\right).
\end{align*}
To prove (\ref{v}) we slightly modify the proof of (\ref{iv}) in the following manner.
From (\ref{i}) we notice that for every prime $p$ dividing $s_n$,
the probability that it divides the missed part of $s_n$ satisfies:
$$
\mathcal{P}\left(p\mid \frac{s_n}{\tilde{s}_n}\right)\leq \left(\frac{1}{2}\right)^r.
$$
As there are at most $\log \left(H\right)$ such primes, we get
$$
\mathcal{P}\left(s_n=\tilde{s}_n\right) \geq 1-\log \left(H\right) \left(1/2\right)^{r} \geq 1-\log \left(H\right)
2^{-\log \left(\log \left(H\right)\right)- \log\left(\frac{1}{\epsilon}\right)} =1-\log \left(H\right) \frac{1}{\log \left(H\right)} \epsilon.
$$
\end{proof}
\begin{remark}\label{rmk:lif}
Theorem \ref{thm:lif} enables us to produce a LIF procedure, which
gives an output $\tilde{s}$ close to $s_n$ with the time complexity
$O\left(n^3\left(\log\left(n\right)+\log\left(\|A\|\right)\right)^2\right)$
(see \eqref{ii}).
\end{remark}

\subsection{Abbott-Bronstein-Mulders,
  Saunders-Wan and Eberly-Giesbrecht-Villard ideas}
Now, the idea of \cite{Abbott1999} is to combine both
the Chinese remainder and the LIF approach.
Indeed, one can first compute $s_n$ and then reconstruct only
the remaining factors of the determinant by reconstructing
$\det\left(A\right)/s_n$. The expected complexity of
this algorithm is $O\left(n^{\omega}\log\left(|\det(A)/s_n|\right) + n^3\log^2(n\|A\|)\right)$
which is unfortunately $O^{\sim}\left(n^{\omega+1}\right)$ in the worst case.

Now Saunders and Wan \cite{Saunders2004,Wan2005} proposed a way to
compute not only $s_n$ but also $s_{n-1}$ (which they call a bonus)
in order to reduce the size of the remaining factors $\det(A)/\left(s_n s_{n-1}\right)$. The complexity doesn't
change.

Then, Eberly, Giesbrecht and Villard have shown that for the dense
case the expected number of non trivial invariant factors is small, namely less than
$ 3 \lceil \log_{\lambda}\left(n\right) \rceil + 29$ if the entries of the matrix are
chosen uniformly and randomly in a set of $\lambda$ consecutive integers \cite{Eberly2000}.
As they also give a way to compute any $s_i$, this leads to an
algorithm with the expected complexity
$O\left( n^3 \log^2(n\|A\|) \log\left(n\right)\log_{\lambda}\left(n\right)\right)$.

Our analysis yields that the bound on the expected number of invariant factors for

a random dense matrix can be refined as $O\left(\log^{0.5}\left(n\right)\right)$.

Then our idea is to extend the method of Saunders and Wan to get the last
 invariant factors of $A$ slightly faster than by
\cite{Eberly2000}. 
Moreover, we will show in the
following sections that we are able to build an adaptive algorithm
solving a minimal number of systems.

The analysis also yields that it should be possible to change a $\log
\left(n\right) $ factor in the expected complexity of \cite{Eberly2000}
to a $\log (\log \left(n\right))$. This would require a small modification in the
algorithm and a careful analysis. Assuming that the number of invariant factors is
the expected i.e. it equals $N=O\left(\log\left(n\right)\right)$, we can verify the hypothesis by computing
the $(n-N-1)$th factor. If it is trivial, the binary search is done among
$O\left(\log\left(n\right)\right)$ elements and there are only $O\left(\log\left(n\right)\right)$ factors to
compute, which allows to lessen the probability of correctness of each
OIF procedure. Thus, in the propitious case, the expected complexity of the
algorithm would be
$O\left(n^3\log^2(n\|A\|)\log_\lambda\left(n\right)\log^2(\log(n))\right)$.
However, this cannot ce considered as the average complexity in the
ordinary sense since we do not average over all possible inputs in
the analysis.

\section{Computing the product of $O\left(\log(n)\right)$ last invariant factors}\label{sec:last}
\subsection{On the number of invariant factors}\label{number}
The result in \cite{Eberly2000} says  that a $n\times n$ matrix with
entries chosen randomly and uniformly from a set of size $\lambda$ has
the expected number of invariant factors bounded by $3\lceil \log_{\lambda}\left(n\right)\rceil+29$. In
search for some sharpening of this result we prove the following theorems.
\begin{theorem}\label{modular}
Let $A$ be an $n\times n$ matrix with
entries chosen randomly and uniformly from the set of contiguous
integers $\{-\lfloor \frac{\lambda}{2}\rfloor \dots \lceil \frac{\lambda}{2}\rceil\}$. Let $p$ be a prime. The expected number of non-trivial invariant
factors of $A$ divisible by $p$ is at most 4.
\end{theorem}
\begin{theorem}\label{expected}
Let $A$ be an $n\times n$ matrix with
entries chosen randomly and uniformly from the set $\{-\lfloor \frac{\lambda}{2}\rfloor \dots \lceil \frac{\lambda}{2}\rceil\}$.
The expected number of non trivial invariant factors of $A$ is at most
$\left\lceil\sqrt{2\log_{\lambda}\left(n\right)}\right\rceil+3$.
\end{theorem}

In order to prove the theorems stated
above, we start with the following lemmas.
\begin{lemma}\label{app:lemma}
If $j\hspace{-3pt}>$1
the sum $\hspace{-5pt}\displaystyle\sum_{8 < p < \lambda}\hspace{-5pt} \left(\frac{1}{\lambda}\lceil
\frac{\lambda}{p}\rceil\right)^{j}\hspace{-5pt}$ over primes $p$ can be upper bounded by $\left(\frac{1}{2}\right)^j$.
\end{lemma}
\begin{proof}
We will consider separately the primes from the interval
$\frac{\lambda}{2^{k+1}} \leq p < \frac{\lambda}{2^k}$,
$k=0,1,\dots k_{max}$. The value of $k_{max}$ is computed from the
condition $p>8$ and is equal to $\lceil\log\left(\lambda\right)\rceil-4$. For the $k$th
interval $\lceil\frac{\lambda}{p}\rceil$ is less than or equal to $2^{k+1}$. In each interval
there are at most $\lceil\frac{\lambda}{2^{k+2}}\rceil$ odd numbers and at
most $\frac{\lambda}{2^{k+2}}$ primes: if in the interval there are
more than 3 odd numbers,
at least one of them is divisible by $3$ and is therefore composite.
For this to happen it is enough that 
$\lceil\frac{\lambda}{2^{k_{max}+2}}\rceil\geq 3$, which is the case.
We may therefore calculate:
\begin{align*}
&\sum_{8 < p < \lambda}\left(\frac{1}{\lambda}\lceil\frac{\lambda}{p}\rceil\right)^j \hspace{-3pt}\leq\hspace{-3pt}
\sum_{k=0}^{\lceil\log\left(\lambda\right)\rceil-4}\hspace{-5pt}\frac{\lambda}{2^{k+2}}\left(\frac{2^{k+1}}{\lambda}\right)^{j}\hspace{-3pt}\leq\hspace{-3pt}
\sum_{k=0}^{\lceil\log\left(\lambda\right)\rceil-4}\hspace{-2pt}\frac{1}{2}\left(\frac{2^{k+1}}{\lambda}\right)^{j-1}\hspace{-12pt}=
\frac{1}{2\lambda^{j-1}}\hspace{-8pt}\sum_{k=0}^{\lceil\log\left(\lambda\right)\rceil-4}\hspace{-12pt}\left(2^{k+1}\right)^{j-1}\\&\leq
\frac{1}{2\lambda^{j-1}}\left(\sum_{k=0}^{\lceil\log\left(\lambda\right)\rceil-4}2^{k+1}\right)^{j-1}\hspace{-15pt}\leq
\frac{1}{2\lambda^{j-1}}\left(2^{\lceil\log\left(\lambda\right)\rceil-2}\right)^{j-1}\hspace{-10pt}\leq
\frac{1}{2\lambda^{j-1}}\left(2^{\log\left(\lambda\right)-1}\right)^{j-1}\hspace{-13pt}=\left(\frac{1}{2}\right)^{j}\hspace{-5pt}.
\end{align*}
\end{proof}
\begin{remark}
For $\lambda=2^{l}$, $k$ can be allowed from $0$ up to
$l-3$, instead of $\lceil\log\left(\lambda\right)\rceil-4$ and we can include
more primes in the sum. As a result we obtain an inequality
$\sum_{4 < p < 2^l} \left(\frac{1}{2^l}\lceil
\frac{2^l}{p}\rceil\right)^{j} \leq \left(\frac{1}{2}\right)^j$.
\end{remark}

\begin{lemma}\label{lem:rankp}
Let $A$ be a $k\times n$, $k \leq n$ integer matrix with entries
chosen uniformly and randomly from the set $\{-\lfloor \frac{\lambda}{2}\rfloor \dots \lceil \frac{\lambda}{2}\rceil\}$ . The probability
that $\rank_p(A)$, the rank modulo $p$ of $A$, is $j$, $0 < j \leq
k$ is less than or equal to 
\begin{align}\label{eq:bound}\nonumber
\mathcal{P}\left(\rank_p(A)= j \right)&\leq
\prod_{i=0}^{j-1}(1-\alpha^{(n-i)})\cdot \beta^{(n-j)(k-j)}\cdot
\left(\frac{1}{1-\beta}\right)^{\max{k-j-1,0}}(1+\beta\dots\beta^{k-j})
\\&\leq \beta^{(n-j)(k-j)}\left(\frac{1}{1-\beta}\right)^{k-j},
\end{align}
where $\alpha=\frac{1}{\lambda+1}\lfloor \frac{\lambda+1}{p}\rfloor$ and $\beta=\frac{1}{\lambda+1}\lfloor \frac{\lambda+1}{p}\rfloor$.
\end{lemma}

The proof of the lemma is given in the appendix \ref{app:rankp}.

\begin{proof} {(Theorem \ref{modular})}
\\
The idea of the proof is similar to that of \cite{Eberly2000}[Thm. 6.2].

For $k\geq j$ let $\MDep_k\left(p,j\right)$ denote the event that the first $k$ columns of $A$ mod
$p$ have rank at most $k-j$ over $\Z_p$. 
By $I_j\left(p\right)$ we denote the event, that at least $j$ invariant factors
of $A$ are divisible by $p$. This implies that the first columns of $A$ have
rank at most $n-j$ mod $p$, or that $\MDep_{n-j+k}\left(p,k\right)$ has occurred
for all $k=0\dots j$. This proves in particular that
$\mathcal{P}\left( I_j\left(p\right) \right) \leq
\mathcal{P}\left(\MDep_{n}\left(p,j\right)\right)$.

In order to compute the probability $\mathcal{P}\left(\MDep_{s}\left(p,j\right)\right)$, $s\geq j$ we
notice that it is less than or equal to 
$$
\mathcal{P}\left(\MDep_{s}\left(p,j\right)\right)\leq
\mathcal{P}\left(\MDep_{j}\left(p,j\right)\right)+\sum_{k=j+1}^{s}\mathcal{P}\left(\MDep_{k}\left(p,j\right)\wedge\neg\MDep_{k-1}\left(p,j\right)\right)
$$

Surely,
$\MDep_j\left(p,j\right)$ means that the first $j$ columns of $A$ are 0 mod $p$,
and consequently the probability is less than or equal to
$\beta_p^{jn}$, where the value $\beta_p=\frac{1}{\lambda+1}\lceil
\frac{\lambda+1}{p}\rceil$ is a bound on the probability that an entry of the
matrix is determined modulo $p$ and is set to $(\lambda+1)^{-1}$ if
$p\geq\lambda+1$ or less than or equal $\frac{2}{p+1}$ in the case $p < \lambda+1$.
%
%

We are now going to find
$\mathcal{P}\left(\MDep_k\left(p,j\right)\wedge \neg \MDep_{k-1}\left(p,j\right)\right)$ for $k>j$. Since the event $\MDep_{k-1}\left(p,j\right)$ did not occur,
$A_{k-1}$ has rank modulo $p$ at least $\left(k-j\right)$ and of course at most $\left(k-1\right)$. For $\MDep_{k,j}$ to occur it must be
exactly $\left(k-j\right)$. This means that we can rewrite $\mathcal{P}\left(\MDep_k\left(p,j\right)\wedge\neg
\MDep_{k-1}\left(p,j\right)\right)$ as 
$$\mathcal{P}\left(\MDep_{k}\left(p,j\right)\mid\rank_p\left(A_{k-1}\right)=k-j\right)\cdot\mathcal{P}\left(\rank_p\left(A_{k-1}\right)=k-j\right),$$
 where
$\rank_p\left(A_{k-1}\right)$ denotes the rank modulo $p$ of submatrix $A_{k-1}$
of $A$, which consists of its first $\left(k-1\right)$ columns.

Since the rank modulo $p$ of $A_{k-1}$ is equal to $k-j$, there exists a set of $k-j$ rows $L_{k-j}$
which has full rank mod p. This means that we can choose $k-j$ entries
of the $k$th column randomly but the remaining $n-k+j$ entries will be
determined modulo $p$. This leads to an inequality 
$$
\mathcal{P}\left(\MDep_k\left(p,j\right)~|~
\rank_p\left(A_{k-1}\right)=k-j\right) \leq \beta_p^{n-k+j}.
$$

By Lemma \ref{lem:rankp} we have $\mathcal{P}(\rank(A_{k-1}=k-j)\leq
\left(\frac{1}{1-\beta_p}\right)^{j-1}\beta_p^{(n-k+j)(j-1)}$. Finally, we get  
\begin{equation}
\mathcal{P}\left(\MDep_k\left(p,j\right)\wedge\neg
\MDep_{k-1}\left(p,j\right)\right)\leq \left(\frac{1}{1-\beta_p}\right)^{j-1}\beta_p^{(n-k+j)j}
\end{equation}
and 
\begin{equation}\label{eq:mdepspj}
\mathcal{P}\left(\MDep_{s}\left(p,j\right)\right)\leq \left(\frac{1}{1-\beta_p}\right)^{j-1}\sum_{k=j}^{s}\beta_p^{\left(n-k+j\right)j}
< \left(\frac{1}{1-\beta_p}\right)^{j-1}\beta_p^{j(n-s+j)}\frac{1}{1-\beta_p^j}.
\end{equation}

The expected number of invariant factor
divisible by $p<\lambda$ verifies:
\begin{align*}\sum_{j=0}^n j \left( P\left( I_j\left(p\right) \right) - P\left( I_{j+1}\left(p \right)\right) \right) &=
\sum_{j=1}^n P\left( I_j\left(p\right) \right)
\leq  \sum_{j=1}^n \MDep_{n}\left(p,j\right) \\ &\leq
 \sum_{j=1}^n \left(\frac{p+1}{p-1}\right)^{j-1}\left(\frac{2}{p+1}\right)^{j^2}\hspace{-5pt}\frac{\left(p+1\right)^j}{\left(p+1\right)^j-2^j}
\end{align*}
The latter is decreasing in $p$ and therefore less than its value at
$p=2$, which is lower than 
3.46.

For $p\geq \lambda+1$ the result is even sharper:
$$\sum_{j=0}^n j \left( P\left( I_j\left(p\right) \right) - P\left( I_{j+1}\left(p \right)\right) \right)
\leq
\sum_{j=1}^n
\left(\frac{\lambda}{\lambda-1}\right)^{j-1}
\left(\frac{1}{\lambda}\right)^{j^2}\frac{\lambda^j}{\lambda^j-1}
\leq \frac{1}{\lambda-1}\frac{1}{1-\frac{1}{(\lambda-1)\lambda^2}}
$$
the latter being lower than 1.18 for $\lambda\geq 1$.
\end{proof}

\begin{proof} {(Theorem \ref{expected})}

In addition to $\MDep_k\left(p,j\right)$ introduced earlier, let 
$\Dep_k$ denote an event that the first $k$ columns of $A$ are linearly
dependent (over rationals) and $\MDep_k\left(j\right)$, an event that either of $\MDep_k\left(p,j\right)$ occurred.

Recall from \cite[\S 6]{Eberly2000} that 
\begin{align*}
&\mathcal{P}(\Dep_1 )\leq (\lambda+1)^{-n}\\&
\mathcal{P}\left(\Dep_k \wedge \neg
\Dep_{k-1}\right)\leq\mathcal{P}\left(\Dep_k \mid \neg
\Dep_{k-1}\right)\leq (\lambda+1)^{-n+k-1}.
\end{align*}
This gives $\mathcal{P}(\Dep_k)\leq \frac{1}{(\lambda+1)^n}+\dots+
\frac{1}{(\lambda+1)^{n-k+1}}$ which is less than $\frac{1}{(\lambda+1)^{n-k+1}}\frac{\lambda+1}{\lambda}$.

As in the previous proof, the probability that the number of non trivial invariant
factors is at least
 $j$ (event $I_j$) is lower than
$\mathcal{P}\left( \MDep_{n-j+k}(k) \vee \Dep_{n-j+1}\right) $ for all $k=0\dots j$. 
The latter can be transformed to $\mathcal{P}\left(
(\MDep_{n-j+k}(k)\wedge \neg \Dep_{n-j+1}) \vee \Dep_{n-j+1}\right) $,
and both $\mathcal{P}\left(\MDep_{n-j+k}(k)\wedge \neg
\Dep_{n-j+1}\right)$ and $\mathcal{P}\left(\Dep_{n-j+1}\right)$ can be
treated separately.

To compute $\mathcal{P}\left(\MDep_{n-j+k}(k)\wedge \neg
\Dep_{n-j+1}\right)$ we will sum $\mathcal{P}\left(\MDep_{n-j+k}(p,k)\right)$
over all possible primes. Since $\Dep_{n-j+1}$ does not hold, there
exists a $(n-j+1)\times (n-j+1)$ non-zero minor, and we have to sum
over the primes which divide it. We will treat separately primes $p <
\lambda+1$ and $p\geq \lambda+1$. 
Once again we set
$\beta_p=\frac{2}{p+1}$ for $p < \lambda+1$ and
$\beta_p=\frac{1}{\lambda+1}$ for $p \geq \lambda$.

By \eqref{eq:mdepspj} we have 
\begin{align*}
&\sum_{p<\lambda}\mathcal{P}(\MDep_{n-j+k}(p,k)) < \left(\frac{1}{1-\beta_2}\right)^{k-1}\beta_2^{kj}\frac{1}{1-\beta_2^k}+\left(\frac{1}{1-\beta_3}\right)^{k-1}\beta_3^{kj}\frac{1}{1-\beta_3^k}\\+
&\left(\frac{1}{1-\beta_5}\right)^{k-1}\beta_5^{kj}\frac{1}{1-\beta_5^k}+
\left(\frac{1}{1-\beta_7}\right)^{k-1}\beta_7^{kj}\frac{1}{1-\beta_7^k}+
\hspace{-4pt}\sum_{8<p<\lambda}\left(\frac{1}{1-\beta_p}\right)^{k-1}\beta_p^{kj)}\frac{1}{1-\beta_p^k}.
\end{align*}
This transforms to 
\begin{align*}
&\sum_{p<\lambda}\mathcal{P}(\MDep_{n-j+k}(p,k))\leq
3^{k-1}\left(\frac{2}{3}\right)^{kj}\frac{3^k}{3^k-2^k}+
2^{k-1}\left(\frac{1}{2}\right)^{kj}\frac{2^k}{2^k-1}\\+
&\left(\frac{3}{2}\right)^{k-1}\left(\frac{1}{3}\right)^{kj}\frac{3^k}{3^k-1}+
\left(\frac{4}{3}\right)^{k-1}\left(\frac{1}{4}\right)^{kj}\frac{4^k}{4^k-1}\\&+
\left(\frac{6}{5}\right)^{k-1}\frac{6^k}{6^k-1}\sum_{8<p<\lambda+1}\left(\frac{1}{\lambda+1}\lceil\frac{\lambda+1}{p}\rceil\right)^{kj}.
\end{align*}
Thanks to Lemma \ref{app:lemma}, the sum 
$\sum_{8<p<\lambda+1}\left(\frac{1}{\lambda+1}\lceil\frac{\lambda+1}{p}\rceil\right)^{kj}$
can be bounded by $\left(\frac{1}{2}\right)^{kj}$. 

For primes $p \geq \lambda+1$ we should estimate the number of
primes dividing the $(n-j+1)$th minor. By the Hadamard's bound (notice that
$\Dep_{n-j+1}$ does not hold), the minors are bounded in absolute value by
$\left(\left(n-j+1\right)\left(\frac{\lambda+1}{2})\right)^2\right)^{\frac{n-j+1}{2}}$. Therefore the number of primes $p\geq
\lambda+1$ dividing the minor is at most
$\frac{n}{2}\left(\log_{\lambda+1}(n) + 2\right)$. 
Summarizing,
\begin{align*}
&\mathcal{P}\left(\left(\MDep_{n-j+k}\left(k\right)\wedge
\neg\Dep_{n-j+1}\right) \vee \Dep_{n-j+1}\right)\leq
\left(\frac{2}{3}\right)^{kj}\frac{3^{2k-1}}{3^k-2^k}+
\left(\frac{1}{2}\right)^{kj}\frac{2^{2k-1}}{2^k-1}\\+
&\left(\frac{3}{2}\right)^{k-1}\left(\frac{1}{3}\right)^{kj}\frac{3^k}{3^k-1}+
\left(\frac{4}{3}\right)^{k-1}\left(\frac{1}{4}\right)^{kj}\frac{4^k}{4^k-1}+
\left(\frac{6}{5}\right)^{k-1}\frac{6^k}{6^k-1}\left(\frac{1}{2}\right)^{kj}+\\
&+\frac{n}{2}\left(\log_{\lambda+1}(n)+2\right)\left(\frac{\lambda+1}{\lambda}\right)^{k-1}\frac{1}{(\lambda+1)^{jk}}\frac{(\lambda+1)^k}{(\lambda+1)^k-1}
+\frac{\lambda}{\lambda-1}\lambda^{-(n-j+1)}.
\end{align*}

We can now compute the expected number of non trivial invariant
factors.

Let us fix $h = \max (2,\left\lceil \sqrt{2\log_{\lambda+1}\left(n\right)}\right\rceil
)$. We have that in particular, $\mathcal{P}(I_j)$ is less than $\mathcal{P}\big((\MDep_{n-j+h}(h)\wedge\neg\Dep_{n-j+1})\vee\Dep_{n-j+1}\big)$.
We can check that $h^2 \geq  \log_{\lambda+1}\left(n\right)
+  \log_{\lambda+1}\left(\log_{\lambda+1}\left(n\right)+2\right)$.
This gives also
$(\lambda+1)^{h^2} > n\left(\log_{\lambda+1}\left(n\right)+2\right)$ and
$$1 >
\frac{n}{2}\left(\log_{\lambda+1}\left(n\right)+2\right)\left(\frac{\lambda+1}{\lambda}\right)^{h-1}\frac{(\lambda+1)^{2h}}{((\lambda+1)^h-1)^2}\frac{1}{(\lambda+1)^{h(h+1)}}.$$
The expected number of non trivial invariant factors
is bounded by:
$$\sum_{j=1}^h 1 + \hspace{-10pt}\sum_{j=h+1}^n
\mathcal{P}\left((\MDep_{n-j+h}\left(h\right)\vee \neg\Dep_{n-j+1})\wedge \Dep_{n-j+1}\right)$$
which in turn is bounded by
\begin{align*}
&h + \Big(\sum_{j=h+1}^{n} \left(\frac{2}{3}\right)^{hj}\frac{3^{2h-1}}{3^h-2^h}+
\left(\frac{1}{2}\right)^{hj}\frac{2^{2h-1}}{2^h-1}+
\left(\frac{3}{2}\right)^{h-1}\left(\frac{1}{3}\right)^{hj}\frac{3^h}{3^h-1}\\&+
\left(\frac{4}{3}\right)^{h-1}\left(\frac{1}{4}\right)^{hj}\frac{4^h}{4^h-1}+
\left(\frac{6}{5}\right)^{h-1}\frac{6^h}{6^h-1}\left(\frac{1}{2}\right)^{hj}\\&
+\frac{n}{2}\left(\log_{\lambda+1}(n)+2\right)\left(\frac{\lambda+1}{\lambda}\right)^{h-1}\frac{1}{(\lambda+1)^{hj}}\frac{(\lambda+1)^h}{(\lambda+1)^h-1}
+\frac{(\lambda+1)}{\lambda}(\lambda+1)^{-(n-j+1)}
\Big)\\
&\leq
h +    \frac{3^{3h-1}}{(3^h-2^h)^2}\left(\frac{2}{3}\right)^{h(h+1)}
+      \frac{2^{3h-1}}{(2^h-1)^2}  \left(\frac{1}{2}\right)^{h(h+1)}
+\left(\frac{3}{2}\right)^{h-1}\frac{3^{2h}}{(3^h-1)^2}\left(\frac{1}{3}\right)^{h(h+1)}\\
&+\left(\frac{4}{3}\right)^{h-1}\frac{4^{2h}}{(4^h-1)^2}\left(\frac{1}{4}\right)^{h(h+1)}
+\left(\frac{6}{5}\right)^{h-1}\frac{6^h}{6^h-1}\frac{2^{h}}{2^h-1}\left(\frac{1}{2}\right)^{h(h+1)}\\
&+\frac{n}{2}\left(\log_{\lambda+1}(n)+2\right)\left(\frac{\lambda+1}{\lambda}\right)^{h-1}\frac{(\lambda+1)^{2h}}{((\lambda+1)^h-1)^2}\frac{1}{(\lambda+1)^{h(h+1)}}
+\left(\frac{\lambda+1}{\lambda}\right)^2\frac{1}{\lambda+1}\\
&\leq h + f(n,\lambda) +
\frac{n}{2}\left(\log_{\lambda+1}\left(n\right)+2\right)\left(\frac{\lambda+1}{\lambda}\right)^{h-1}\frac{(\lambda+1)^{2h}}{((\lambda+1)^h-1)^2}\frac{1}{(\lambda+1)^{h(h+1)}}
\\&\leq h + f(n,\lambda) + 1.
\end{align*}

where $f\left(n,\lambda\right) \leq
\frac{3^{3h-1}}{(3^h-2^h)^2}\left(\frac{2}{3}\right)^{h(h+1)}
+      \frac{2^{3h-1}}{(2^h-1)^2}  \left(\frac{1}{2}\right)^{h(h+1)}
+\left(\frac{3}{2}\right)^{h-1}\frac{3^{2h}}{(3^h-1)^2}\left(\frac{1}{3}\right)^{h(h+1)}\\
+\left(\frac{4}{3}\right)^{h-1}\frac{4^{2h}}{(4^h-1)^2}\left(\frac{1}{4}\right)^{h(h+1)}
+\left(\frac{6}{5}\right)^{h-1}\frac{6^h}{6^h-1}\frac{2^{h}}{2^h-1}\left(\frac{1}{2}\right)^{h(h+1)}+\left(\frac{\lambda+1}{\lambda}\right)^2\frac{1}{\lambda+1}
< 2 $ 
as soon as $\lambda \geq 2$. On the other hand
$f(n,1)=\left(\frac{2}{2-1}\right)^2\frac{1}{2}\leq 2$ which leads to
the final result.

\end{proof}

\subsection{Extended Bonus Ideas}\label{sec:extended}
In his thesis \cite{Wan2005}, Z. Wan introduces the idea of computing
the penultimate invariant factor (i.e. $s_{n-1}$) of
$A$ while computing $s_{n}$ using two system solvings. The
additional cost is comparatively small, therefore $s_{n-1}$ is
referred to as a bonus. Here, we extend this idea to the computation of the
$\left(n-k+1\right)$th factor with $k$ solvings in the following manner:

\begin{enumerate}
 \item The (matrix) solution of $AX=B$, where $B$ is a
  $n \times k$ multiple right hand side can be written as
$\tilde{s}^{-1}_nN$ where $\tilde{s}_n$ approximates $s_n\left(A\right)$ and
the factors of $N$ give some divisors of the last $k$ invariant factors
of $A$: see lemma \ref{lem:sf}.
\item We are actually only interested in getting the product of
  these invariant factors which we compute as the $\gcd$ of the
determinants of two perturbed $k\times k$ matrix $R_1N$ and $R_2N$. 
\item Then we show that repeating this solving twice with two distinct
  right-hand sides $B_1$ and $B_2$ is in general sufficient to remove
  those extra factors and to get a very fine approximation of the
  actual product of the last $k$ invariants: see lemma \ref{lem:pi}.
\end{enumerate}

\subsubsection{The last $k$ invariant factors}
Let $X$ be a (matrix) rational solution of the equation $AX=B$, where
$B=[b_{i}], i=1,\dots, k$, is a random $n \times k$ matrix. Then the coordinates of
$X$ have a common denominator $\tilde{s}_{n}$ and we let $N=[n_{i}],
i=1,\dots, k$, denote the matrix of numerators of $X$. Thus,
$X=\tilde{s}^{-1}_nN$ and  $\gcd\left(N_{ij},\tilde{s}_{n}\right)=1$.

Following Wan, we notice that $s_{n}\left(A\right)A^{-1}$ is
an integer matrix, the Smith form of which is equal to
$$
diag\left(\frac{s_n\left(A\right)}{s_n\left(A\right)},\frac{s_n\left(A\right)}{s_{n-1}\left(A\right)},\dots,\frac{s_n\left(A\right)}{s_1\left(A\right)}\right).
$$
Therefore, we may compute $s_{n-k+1}\left(A\right)$ when knowing
$s_{k}\left(s_{n}\left(A\right)A^{-1}\right)$.
The trick is that the computation of $A^{-1}$ is not required:
we can perturb $A^{-1}$ by right multiplying it by $B$. Then, $s_{k}\left(s_{n}\left(A\right)A^{-1}B\right)$ is a multiple of
$s_{k}\left(s_{n}\left(A\right)A^{-1}\right)$. Instead of $s_{n}\left(A\right)A^{-1}B$ we
would prefer to use $\tilde{s}_{n}A^{-1}B$ which is
already computed and equal to $N$.

The relation between $A$ and $N$ is as follows.
\begin{lemma}\label{lem:sf}
Let $X=\tilde{s}_n^{-1}N$, $\gcd\left(\tilde{s}_n,N\right)=1$ be a solution to the
equation $AX=B$, where $B$ is $n\times k$ matrix. 
Let $R$ be a random  $k\times n$ matrix.
Then
$$
\left.\frac{\tilde{s}_n}{\gcd\left(s_{i}\left(N\right),\tilde{s}_n\right)}\right| s_{n-i+1}\left(A\right)
~\text{and}~\left.\frac{\tilde{s}_n}{\gcd\left(s_{i}\left(RN\right),\tilde{s}_n\right)}\right| s_{n-i+1}\left(A\right),
\hfill i=1\dots ,k.
$$
\end{lemma}
\begin{proof}
The Smith forms of
$s_{n}\left(A\right)A^{-1}B$ and $N$ are connected by the
relation
$\frac{s_n\left(A\right)}{\tilde{s}_n}s_i\left(N\right)$ $=s_i\left(s_{n}\left(A\right)A^{-1}B\right)$,
$i=1,\dots,k$. Moreover, $s_i\left(N\right)$ is a factor of $s_i\left(RN\right)$.
We notice that
$\frac{s_n\left(A\right)}{s_{i}\left(s_{n}\left(A\right)A^{-1}B\right)}$ equals
$\frac{\tilde{s}_n}{s_{i}\left(N\right)}$,
and thus $\frac{\tilde{s}_n}{\gcd\left(s_{i}\left(RN\right),
\tilde{s}_n\right)}$ is an (integer) factor of $s_{n-i+1}\left(A\right)$. Moreover, the
under-approximation is solely due to the choice of $B$ and $R$.
\end{proof}
\begin{remark}\label{rmk:small_primes}
Taking $\gcd\left(s_{i}\left(RN\right), \tilde{s}_n\right)$ is necessary as $\frac{\tilde{s}_n}{s_{i}\left(RN\right)}$ may
be a rational number.
\end{remark}

\subsubsection{Removing the undesired factors}
In fact we are interested in computing the product
$\pi_k=s_{n}s_{n-1}\cdots s_{n-k+1}\left(A\right)$ of the $k$ biggest invariant factors of
$A$. Then, following the idea of \cite{Abbott1999}, we would like to reduce the
computation of the determinant to the computation of
$\frac{\det\left(A\right)}{\tilde{\pi}_k}$, where $\tilde{\pi}_k$ is a factor of
$\pi_k$ that we have obtained. We can compute $\tilde{\pi}_k$ as
$\tilde{s}_n^{k}/\gcd\left(\mu_k\left(RN\right),
\tilde{s}_n^{k}\right)$, where $\mu_k=s_1 s_2\cdots s_{k}$ is the
product of the $k$ smallest invariant factors.

We will need a following technical lemma. Its proof is given in the
appendix, see \ref{app:determinant}.

\begin{lemma}\label{lem:determinant}
Let $V$ be an $k\times n$ matrix, such that the Smith form of $V$ is
trival. Let $M$ be an $n\times k$ matrix
with entries chosen randomly and uniformly from the set
$\{a,a+1\dots a+S-1\}$, the probability that $p^{l}< S$ divides the
determinant $\det(VM)$ is at most $\frac{3}{p^{l}}$.
\end{lemma}

In the following lemmas we
show that by repeating the choice of matrix $B$ and $R$ twice, we will omit
only a finite number of bits in $\pi_k$. 
We start with a remark, which is a modification of \cite[Lem. 5.17]{Wan2005}. We ramaind that the order modulo $p$ ($\ord_p$) of a value is the expotent of the highest power of $p$ dividing it.
\begin{remark}\label{rmk:ord_p}
For every $n\times n$ matrix $M$ there exist a $k\times n$, $k \leq
n$, matrix $V$ with trivial Smith form, such that
for any $n \times k$ matrix $B$: if the order modulo $p$
$\operatorname{ord}_p\left(\frac{\mu_k\left(MB\right)}{\mu_k\left(M\right)}\right)$
is greater than $l$ then also $\operatorname{ord}_p\left(\det\left(VB\right)\right)$ is
greater than $l$. 
\end{remark}
\begin{lemma}\label{lem:pi}
Let $A$ be an $n\times n $ integer matrix and
$B_i \left(resp. R_i\right)$, $i=1,2$ be $n\times k$ $\left(resp. k\times
n\right)$, matrices with the entries uniformly and
randomly chosen from the set $\mathcal{S}$ of $S$ contiguous integers, $k\geq 2$. Denote by $\mu_k$
the product $s_1\dots s_k$ of the $k$ smallest invariant factors and by
$\pi_k$ the product of the $k$ biggest factors of $A$.
Then for $M=s_n\left(A\right)A^{-1}$
$$
\mathbf{E}\left(\log
\left(
\frac	{\pi_k\left(A\right)}
	{s_n\left(A\right)^{k}}
		\gcd\left( \mu_k(R_1MB_1),\mu_k(R_2MB_2), s_n(A)^k\right)	
\right)\right)\hspace{-3pt}
\in~O\left(1\right)+
O\left(\frac{k^3\log^4\left(H\right)}{S}\right)
$$
where $H$ is the Hadamard bound for $A$.
\end{lemma}
\begin{proof}
First, notice that $\frac{\pi_k\left(A\right)}{s_n\left(A\right)^k}=\frac{1}{\mu_k(M)}$. Therefore

the expected value is less than or equal
\begin{align*}
&\sum_l\sum_{p\arrowvert s_n\left(A\right)}\log\left(p\right)l\mathcal{P}\left(\operatorname{ord}_p\left(\frac{\gcd\left(
\mu_k(R_1MB_1),\mu_k(R_2MB_2),s_n(A)^k\right)}{\mu_k(M)}\right)=l\right)\\=
&\sum_l\sum_{p\arrowvert
s_n(A)}\log(p)\mathcal{P}\left(\operatorname{ord}_p\left(\frac{\gcd\left( \mu_k(R_1MB_1),\mu_k(R_2MB_2),s_n(A)^k\right)}{\mu_k(M)}\right)\geq
l\right)\\\leq
&\sum_l\sum_{p\arrowvert s_n(A)}\log(p)\Pi_{i=1,2}\mathcal{P}\left(\operatorname{ord}_p\left(\frac{ \mu_k(R_iMB_i)}{\mu_k(M)}\right)\geq l\right)\\\leq
&\sum_l\sum_{p\arrowvert s_n(A)}\log(p)\Pi_{i=1,2}\left(\sum_{k=0}^l\mathcal{P}\left(\begin{array}{l}\operatorname{ord}_p\left(\frac{ \mu_k(MB_i)}{\mu_k(M)}\right)\geq k~\wedge\\ \operatorname{ord}_p\left(\frac{ \mu_k(R_iMB_i)}{\mu_k(MB_i)}\right)\geq (l-k) \end{array}\right)\right).
\end{align*}
Thanks to remark \ref{rmk:ord_p} we can link this probability to the
probability that $p^l$ divides the determinant of $VB_i$ or $R_iU$,
for matrices $V,U$ which have a trivial Smith form. 

We only consider $p
\arrowvert s_n\left(A\right)$.

For $p^l < S$ Lemma \ref{lem:determinant} gives
\begin{align*}
&\sum_{k=0}^l\mathcal{P}\left(\operatorname{ord}_p\left(\frac{
\mu_k(MB_i)}{\mu_k(M)}\right)\geq k \wedge
\operatorname{ord}_p\left(\frac{
\mu_k(R_iMB_i)}{\mu_k(MB_i}\right)\geq (l-k) \right)\leq\\
&\sum_{k=0}^l \mathcal{P}\left(B_i:\operatorname{ord}_p\left(\det\left(VB_i\right)\right)\geq k \right) \mathcal{P}\left(R_i:
\operatorname{ord}_p\left(\det\left(R_iU\right)\right)\geq k \right)  \leq \left(l+1\right)\frac{3}{p^l}.
\end{align*}

Now the expected size of the under-estimation
is less than or equal to
\begin{align*}
&\log\left(2\right)\left(3+\sum_{l=4}^{\infty}\left(\left(l+1\right)^2\frac{3}{2^{l}}\right)^{2}\right)+
\log\left(3\right)\left(2+\sum_{l=3}^{\infty}\left(\left(l+1\right)\frac{3}{3^{l}}\right)^{2}\right)\\
&+\log\left(5\right)\left(1+\sum_{l=2}^{\infty}\left(\frac{3}{5^{l}}\right)^{2}\right)+
\log\left(7\right)\left(\sum_{l=2}^{\infty}\left(\frac{3}{7^{l}}\right)^{2}\right)
+\sum_{5 < p \leq
H}\sum_{l=1}^{\infty}\log\left(p\right)\left(\frac{3}{p^{l}}\right)^{2}
\\&\leq 4.36 + 2.24 + 1.14 + 0.77 +\sum_{5 < p \leq
H}\hspace{-8pt}\log\left(p\right)\frac{-27p^2+36p^4+9}{\left(p-1\right)^3\left(p+1\right)^3}
\\&\leq 8.51 +
\int_{10}^{\infty}\hspace{-2pt}\log\left(x\right)\frac{-27x^2+36x^4+9}{\left(x-1\right)^3\left(x+1\right)^3}dx\leq
8.51+11.97
\end{align*}
which is $O\left(1\right)$.

For $p^l \geq S$  the probability $\mathcal{P}\left(p^l
| \det\left(M\right) \right)$ is less than $\mathcal{P}\left(p^{\lfloor
\log_p\left(S\right)\rfloor}\arrowvert \det\left(M\right)\right)$ and consequently can be bounded by
$3\min\left(\frac{1}{p},\frac{p}{S}\right)$ which is less than
$\frac{3}{\sqrt{S}}$. The expected size of the underestimation is
\begin{align*}
&\sum_{p \arrowvert s_n\left(A\right)}\sum_{l=\lceil
\log_p\left(S\right)\rceil}^{k\log_p\left(H\right)} \left(l+1\right)^2\log\left(p\right)\left(\frac{3}{\sqrt{S}}\right)^2
\leq 
\sum_{p \arrowvert
s_n\left(A\right)}\hspace{-10pt}\frac{9\log\left(p\right)}{S}\big(\frac{13}{6}k\log_p\left(H\right)+\frac{3}{2}\left(k\log_p\left(H\right)\right)^2\\
&+\frac{1}{3}\left(k\log_p\left(H\right)\right)^3\big)
\leq
k\log^2\left(H\right)\frac{9}{S}\left(\frac{13}{6}+\frac{3}{2}k\log\left(H\right)+\frac{1}{3}k^2\log^2\left(H\right)\right)\leq
\frac{13k^3\log^4\left(H\right)}{S}.
\end{align*}
This is $O\left(\frac{k^3\log^4\left(H\right)}{S}\right)$, which gives the result.
\end{proof}

Another method to compute the product $\mu_k$ of some first invariant
factors of a rectangular matrix $N$ would be to compute several minors
of the matrix and to take the $\gcd$ of them. In our scheme we can
therefore get rid of matrix $R$ which would enable us to use a smaller
bound on $S=O\left(k\log\left(H\right)\right)$ and still preserve a small error of estimation due to the
choice of $B$. However, it is difficult to judge the impact of
choosing only a few minors (instead of all). An experimental
evaluation whether for random $A$ and random $B$ the minors of $N$  
are sufficiently ''randomly'' distributed remains to be done. 

\section{Introspective Algorithm}\label{sec:algorithm}
Now we should incorporate Algorithm \ref{alg:cra} and the ideas presented
in sections  \ref{sec:lif} and \ref{sec:extended} in the form of an
introspective algorithm.

Indeed, we give  a recipe for an auto-adaptive program that implements
several algorithms
of diverse space and time complexities for solving a particular
problem. The best path is chosen at run time, from a self-evaluation
of the dynamic behavior (here we use timings) while processing a given
instance of the problem. This kind of auto-adaptation is called
introspective in \cite{jgd:2006:AHA}.
In the following,
CRA loop refers to Algorithm
\ref{alg:cra}, slightly modified to compute $\det\left(A\right)/K$.
If we re-run the CRA loop, we use the already computed  modular
determinants first whenever possible.

Informally, the general idea of the introspective scheme is:
\begin{enumerate}
\item Initialize the already computed factor $K$ of the determinant to $1$;
\item \label{fflas} Run fast FFLAS LU routines 
in the background
  to get several modular determinants $d_i=\det\left(A\right) \mod p_i$.
\item From time to time try to early terminate the Chinese remainder
  reconstruction of $\det\left(A\right)/K$.
\item In parallel or in sequential,
  solve random systems to get the last invariant factors one after the
 other.
\item Update $K$ with these factors and loop back to step \eqref{fflas} until an early
  termination occurs or until the overall
  timing shows that the expected complexity is exceeded.
\item In the latter exceptional case, switch to a better worst case
  complexity algorithm.
\end{enumerate}
More precisely, the full algorithm in shown on page \pageref{alg:introsp}.
\begin{algorithm}\caption{Extended Bonus Determinant Algorithm}\label{alg:introsp}
\begin{algorithmic}[1]
\REQUIRE An integer $n \times n$ matrix $A$.
\REQUIRE H - bound for $\det(A)$ (can be the Hadamard's bound)
\REQUIRE $0 < \epsilon < 1$, an error tolerance, $S=13\mathbf{E}\left(\#factors\left(A\right)\right)^3\left(\lceil\log\left(H\right)\rceil\right)^4, l > 1$. 
\REQUIRE A stream $\mathcal{S}$ of numbers randomly chosen from the
set of $S$ contiguous integers.
\REQUIRE A set P of random primes greater than $l$, $|P|\geq \lceil
2log_l(H)\rceil$, $P'=|P|-\log_l(H)$
\ENSURE The integer determinant of $A$, correct  with probability at
least $1-\epsilon$.
\smallskip

\STATE \label{eb:k} $k=  \log\left( 1/\epsilon \right)/\lceil\log\left(
\frac{P'}{\log_{l}\left(H\right)} \right)\rceil$;\hfill see Lem. \ref{lem:cra}\eqref{cra:probability}
\FOR{$i=1$ to $k$}\label{eb:ET}
	\STATE run the CRA loop for $\det\left(A\right)$ ;\hfill //see Alg. \ref{alg:cra}
	\IFTHENEND{early terminated}{Return determinant}
\ENDFOR
\STATE $i_{max} = i_{max}\left(A\right), i_{min} = i_{min}\left(A\right)$; \hfill //see \S \ref{rem:imax}
\STATE $\tilde{\pi}_{0}=1;K=1;$
\STATE $k_{done}=0; k_{app}=0; j=0;$
\WHILE {$k_{done} \leq i_{max}$}
\STATE $i=k_{done}+1;$
\WHILE{$i \leq i_{max}$}\label{eb:loop}
	\STATE Generate $b_i^{\left(j\right)}$ a random vector of dimension $n$ from the stream
$S$;
	\STATE \label{eb:lif}Compute $\tilde{s}_n$ by solving
        $Ax_i^{\left(j\right)}=b_i^{\left(j\right)}$; \hfill //see Section \ref{sec:lif}
	\IFTHENELSE{$i=1$}{;$\tilde{\pi}_1=\tilde{s}_n$;}
		\STATE $N := \tilde{s}_n X$, where $X=[x_l^{\left(j\right)}]_{l=1,\dots
i}$;\hfill //see Section \ref{sec:extended};
		\STATE Generate a random $i\times n$ matrix
$R$. 
		\STATE $\tilde{\pi}_{i}=\frac{\tilde{s}_n^{i+1}}{\gcd\left(\det\left(RN\right),\tilde{s}_n^{i+1}\right)}$ \hfill //determinant computation
	\ENDIF
	\STATE $K=\operatorname{lcm}\left(\tilde{\pi}_{i}, K\right)$; $\tilde{\pi}_{i} = K$;
	\STATE \label{eb:underapp}Resume CRA looping on $d=\det\left(A\right)/K$ for at most the time of one system solving;
        \IFTHENEND{early terminated}{Return $d\cdot K$;}
	\IF {$i > i_{min}$}
		\IF{$\tilde{\pi}_{i}=\tilde{\pi}_{i-1}$}
			\IF {$i > k_{app}$}
			\STATE $k_{done}=k_{app}; k_{app} = i; j=j+1 \mod 2;$ break;
			\ELSE
				\STATE Resume CRA looping on $d=\det\left(A\right)/K$ for at most the time of $\left(i_{max}-i\right)$ system solvings;
				\IFTHEN{early terminated}{Return $d\cdot K$;}
			        \ELSEEND{$i=i_{max}$;}
			\ENDIF
		\ENDIF
        \ENDIF
	\STATE i=i+1;
\ENDWHILE
\ENDWHILE
\STATE \label{eb:last}run an asymptotically better
integer determinant algorithm;
\end{algorithmic}
\end{algorithm}
\subsection{Introspectiveness: dynamic choice of the
  thresholds}\label{rem:imax}

The introspective behavior of algorithm  \ref{alg:introsp} depends
paramountly on the number of system solvings and on the size of the
random entries.

The parameter
$i_{max}$ controls the maximal total number of system solvings
authorized before switching to a best worst-case complexity algorithm.
The choice of $i_{max}$ has to be discussed in terms of the expected number of invariant factors of $A$.

First, depending on the size of the set from which we are sampling the
random right-hand sides,
a minimum number of solvings is required to get a good probability of
correctness. We thus define this to be $i_{min}$.

In the dense case, the
(\ref{ii}) part of theorem \ref{thm:lif} states that $i_{min}=2$ is
sufficient. Part (\ref{iv}) part of theorem \ref{thm:lif} 
prompts us to take $i_{min} = \lceil 2\log\left(\log\left(H\right)\right)\rceil$ if we want
to use a smaller $\beta$.

Then this number $i_{min}$ has also to be augmented if the expected
number of  non trivial invariant factors is higher. We thus set
$$i_{max}=\max\left(i_{min},\mathbf{E}\left(\#factors\left(A\right)\right)\right).$$
In
the dense case $\mathbf{E}\left(\#factors\left(A\right)\right)$ is less than
$\lceil\sqrt{2\log_{\lambda}\left(n\right)}\rceil+3$ as shown in theorem
\ref{expected} . 

Now, random vectors are randomly sampled a set of size $S$.
For a dense matrix $A$ we need 
$S=13\mathbf{E}\left(\#factors\left(A\right)\right)^3\left(\lceil\log\left(H\right)\rceil\right)^4$
to get a good probability of success as shown in theorem
\ref{thm:lif}(\ref{ii}) and lemma \ref{lem:pi}.

Additionally, (see lemma \ref{lem:pi}) we should ensure that $\pi_{k}$
is computed twice using different matrices $B$. We therefore
introduce the variables $k_{done}$ and $k_{app}$ which store
respectively the number of
factors computed at least twice (up to $O(1)$)
or once (thus only approximated).

\subsection{Correctness and complexity}\label{sec:complexity}
\begin{theorem}
Algorithm \ref{alg:introsp} correctly computes the determinant with probability $1-\epsilon$.
\end{theorem}
\begin{proof}
Termination is possible only by the early terminated CRA loop or
by the determinant algorithm used in the last step. The choice of $k$
from theorem
\ref{lem:cra}\eqref{cra:probability} and the choice of the determinant
algorithm from \cite{Kaltofen2004, Storjohann2005} ensures that
$1-\epsilon$ probability is obtained.
\end{proof}
%
%
%
The following theorem gives the complexity of the algorithm.
\begin{theorem}
The expected complexity of Algorithm \ref{alg:introsp} in the case of
a dense matrix
is $$O \left(n^{\omega}\log\left(1/\epsilon\right)+n^{3}\left(\log
n+\log\left(\|A\|\right)\right)^2\log^{0.5}\left(n\right)\right).$$

The worst case complexity depends on the algorithm used in the last step.
\end{theorem}
\begin{proof}
To analyze the complexity of the algorithm we will consider the
complexity of each step.

For a dense matrix $A$, with $k$ defined as in the line 1, the
complexity of initial CRA iterations is $O\left(n^{\omega}\log
\left(1/\epsilon\right)\right)$. The while loop is constructed in this
way that we perform at most $2i_{max}$ (see subsection \ref{rem:imax}
for the bound on $i_{max}$) iterations, where 
$\log\left(\|B\|\right)=O\left(\log(n)\log(\log(\|A\|))\right)$. Therefore the cost is
$O\left(n^3\left(\log(n)+\log(\|A\|)\right)^2\hspace{-2pt}\sqrt{\log(n)}\right)$. Considering the time limit,
this is also the time of all CRA loop iterations. To compute
$\tilde{\pi}_i$ we 

need $ni^{\omega-2}$ bit operations. Then, the computation of the
$i\times i$ determinant of $RN$ by a deterministic algorithm (i.e, deterministic CRA)
costs $O(i^{\omega}(\log(i)+n(\log(n\|R\|\cdot\|A\|\cdot\|B\|))))$ bit operations, which for $i=2,\dots, i_{max}$ with $i_{max}$ being $O\left(\log\left(n\right)\right)$ is
$O^{\sim}\left(n\right)$ and thus negligible.

With the expected number of invariant factors bounded by $i_{max}$
(see Thm.\ref{expected}), it is expected that the algorithm will return the
result before the end of the {\em while} loop, provided that the
under-estimation of $\tilde{\pi}_{i_{max}}$ is not too big. But by
updating $\tilde{s}_n$ $O\left(\log^{0.5}\left(n\right)\right)$ times and updating the product
$\tilde{\pi}_{i_{max}}$ twice, it is expected that the overall
under-estimation will be $O(1)$ (see Theorem \ref{thm:lif} and Lemma
\ref{lem:pi}), thus it is possible to recover it by several
CRA loop iterations.
\end{proof}
For the last step for a dense matrix we
propose the $O^{\sim}\left( n^{3.2} \log\left(\|A\|\right) \right)$ algorithm of Kaltofen
\cite{Kaltofen2005} or $O^{\sim}\left(n^{\omega} \log(\|A\|)\right)$ algorithm of
Storjohann \cite{Storjohann2005}. We refer to \cite{Kaltofen2004}
for a survey on complexity of determinant
algorithms. 
\section{Experiments and Further Adaptivity}\label{sec:experiments}

\subsection{Experimental results}
The described algorithm is implemented in the LinBox exact linear
algebra library \cite{Dumas2002Linbox}. In a preliminary version
$i_{max}$ is set to 2 or 1 and the switch in the last step is not
implemented. This is however enough to evaluate the performance of
the algorithm and to introduce further adaptive innovations.

All experiments were performed on 1.3 GHz Intel Itanium2 processor with 128 GB (196 GB since september 2006) of memory disponsible. 

For a generic case of random dense matrices our observation is
that the bound for the number of invariant factors is quite
crude. Therefore the algorithm \ref{alg:introsp} is constructed in the
way that minimizes the number of system solving to at most twice the
actual number of invariant factors for a given matrix. Under the
assumption that the approximations $\tilde{s}_n$ and $\tilde{\pi}_{i}$
are sufficient, this leads to a quick solution.

\begin{figure}
\centering
\includegraphics[width=300pt]{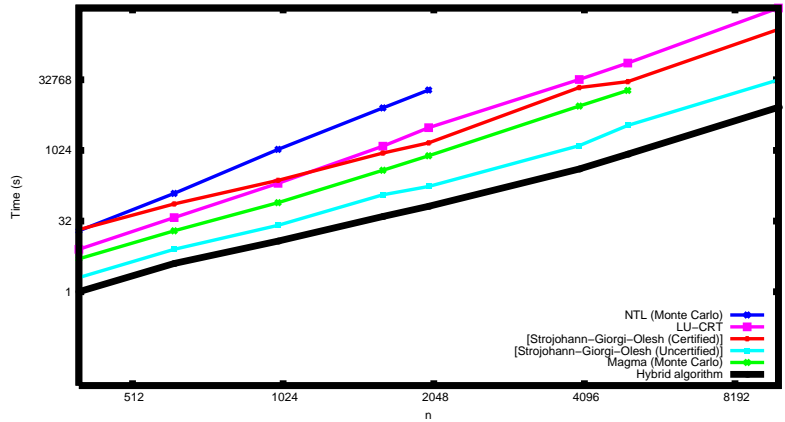}
\caption{Comparison of our algorithm with other existing
implementation. Tested on random dense matrices of the order 400 to
10000, with entries \{-8,-7,\dots,7,8\} Using fast modular routines
puts our algorithm several times ahead of the others. Scaling is logarithmic.}\label{picture}
\end{figure}

\begin{figure}
\centering
\includegraphics[width=300pt]{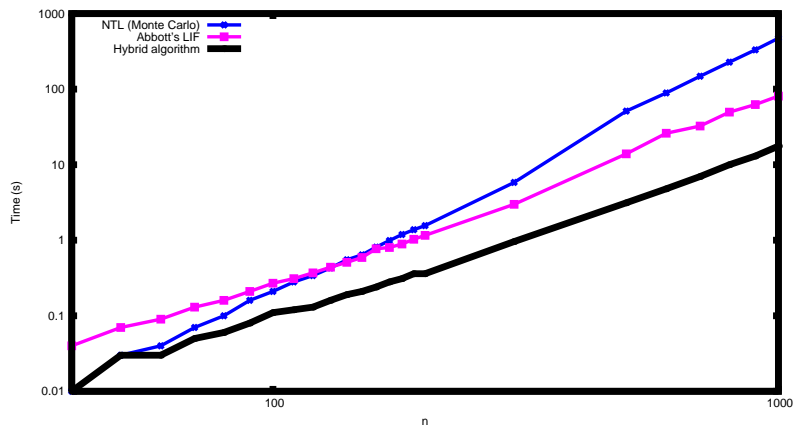}
\caption{Comparison of our algorithm with early terminated Chinese
remaindering algorithm (LU) and the algorithm of Abbott {\em et al.}
\cite{Abbott1999} (LIF). Tested on random dense matrices of the order 40 to
1000, with entries \{-100,-99,\dots,99,100\}. When matrix size exceeds
80 the adaptive algorithm wins. Scaling is logarithmic.}\label{hybrid}
\end{figure}

Indeed for random dense matrices, the algorithm nearly always stopped
with early termination after one system solving.
This together with fast underlying arithmetics of FFLAS \cite{Dumas2002issac} accounted for
the superiority of our algorithm as seen in figure \ref{picture} and \ref{hybrid}
where comparison of timings for different algorithms is
presented. Notice, that our algorithm beats the uncertified 
(i.e. Monte Carlo type)
version of the algorithm of
\cite{Storjohann2004} which claims currently the best theoretical
complexity. This proves that adaptive approach is a powerful tool
which allow us to construct the algorithms very fast in practice 

Thank to the introspective approach our algorithm can detect the
cases when the number of invariant factors is small and equal to $k<
i_{max}$ . One can therefore argue the complexity of our algorithm is in fact $O\left(n^{3}\left(\log\left(n\right) +
\log\left(\|A\|\right)\right)^{2}k\right)$, where $k$
is the number of invariant factors. To test the performance of our
algorithm to detect propitious cases we have run it on various sets of
structured and engineered matrices. The adaptive approach allowed us
to obtain very good timings which motivates us to encourage the use of
this algorithms in the situations which go further beyond the dense
case.

\begin{figure}
\centering
\includegraphics[width=300pt]{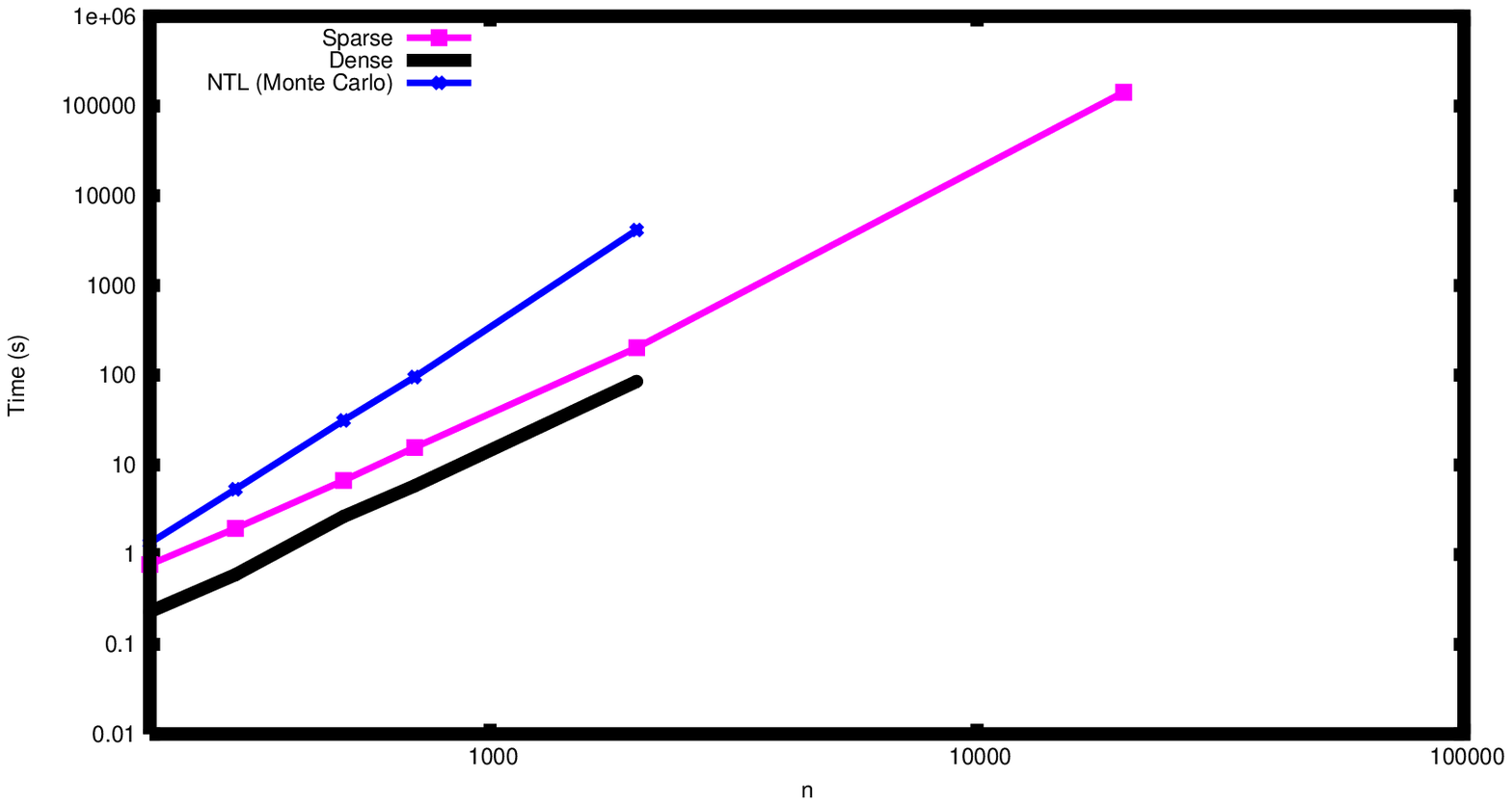}
\caption{Comparison of sparse and dense variants of our determinant algorithm for Trefethen's matrices. Scaling is logarithmic.}\label{sparse}
\end{figure}

Figure \ref{sparse} we present the results of the determinant
computation for sparse matrices of
N. Trefethen\footnote{http://ljk.imag.fr/membres/Jean-Guillaume.Dumas/Matrices/Trefethen/}.

The results encouraged us to construct a sparse variant of our
algorithm, which we shortly describe in Section \ref{sec:sparse}.
Figure \ref{sparse} gives a comparison of the performance of sparse and dense
variants. We used the sparse solver of \cite{EbGies2006}. 
Using the algorithm with the dense solver outperforms using the sparse solver by a factor of $3.3$ to $2.3$,
and decreasing with the matrix size $n$. Thanks to the space-efficiency
of the sparse algorithm we are able to compute the determinant for
$20000\times 20000$ matrix for which the dense solver requires to much memory.

\begin{figure}
\centering
\includegraphics[width=300pt]{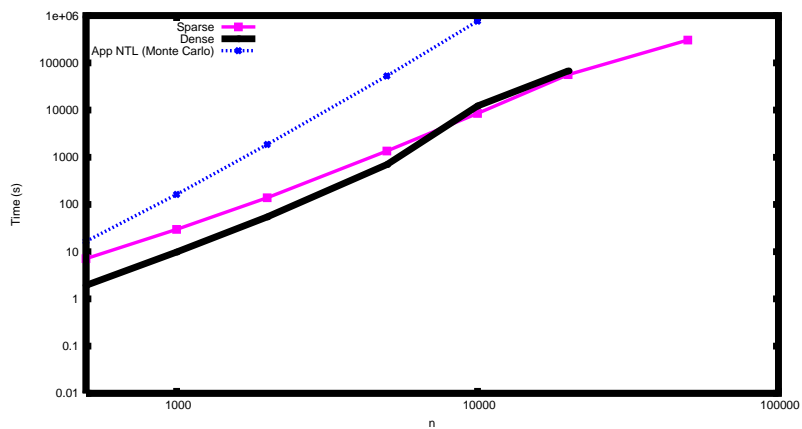}
\caption{Comparison of sparse and dense variants of our determinant
algorithm with the CRA algorithm for random sparse matrices. Scaling is logarithmic. The running
time of the CRA algorithm has been approximated based on the timings
for one iteration}\label{randomsparse}
\end{figure}

In figure \ref{randomsparse} we compare the performance of dense and
sparse variants of the algorithm with the CRA algorithm
(sparse variant) for random sparse matrices. The matrices are very sparse (20 non-zero entries
per row). To ensure that the determinant is non-zero we put 1 on
the diagonal. Both dense and sparse variants of the algorithm have
better running times than the CRA, which proves that we can detect
propitious cases for sparse matrices. Furthermore, sparse variant is
best for bigger matrices and again lets us solve the problem when the
dense variant fails due to unsufficient memory.

In Table \ref{table} we give the timings for the algorithm with
$i_{max}=1$ and $2$. The algorithms was run on a set of specially
engineered matrices which have the
same Smith form as
$diag\{1,2\dots n\}$ and the number of invariant factors of about
$\frac{n}{2}$. We notice that the algorithm
with $i_{max}=1$ (which is in fact a slightly modified version of
Abbott's algorithm \cite{Abbott1999}) runs better for small $n$. This
motivated us to develop an even more adaptive approach, which we
describe in Section \ref{sec:adaptive}.

\begin{table}
\begin{center}
\begin{tabular}{|l|c|c||l|c|c|}
\hline
n	&$i_{max}=1$&$i_{max}=2$&n	&$i_{max}=1$&$i_{max}=2$\\\hline
100	&0.17	&0.22&300	&5.65	&5.53\\
120	&0.29	&0.33&350	&9.76	&9.64\\
140	&0.48	&0.55&400	&14.99	&14.50\\
160	&0.73	&0.78&600	&57.21	&54.96\\
180	&1.07	&1.16&800	&154.74	&147.53\\
200	&1.49	&1.51&1000	&328.93	&309.61\\
250	&2.92	&3.00&2000	&3711.26&3442.29\\\hline
\end{tabular}
\caption{Comparison of the performance of Algorithm \ref{alg:introsp}
with $i_{\max}$ set to 1 and 2 on engineered matrices.} \label{table}
\end{center}
\end{table}

\subsection{Sparse matrix case}\label{sec:sparse}

When trying to adapt our determinant algorithm to the sparse case, the
immediate problem is the bound for the expected number of invariant
factors. On can easily notice, that for a matrix with $k$ non-zero entries per
row, chosen uniformly and randomly from the set of $S$ contiguous
integers, the expected number of invariant factors divisible by $p^l$ can be bounded from
below by $n\frac{1}{S}\lfloor \frac{S}{p^l}\rfloor$ and thus is linear
with $n$. Thus, we cannot use the same argument to estimate the
expected number of system solvings as in the dense case.

One solution would be to consider the number of ''big'' invariant
factors, i.e. the number of invariant factors which are bigger than a
certain parameter $C$. The parameter has to be chosen in a way, that
the product of all smaller factors can be computed by modular CRA loop
quicker than another rational solution of a system of equation. We
could exploit here the difference in complexity between system solving
and one modular routine which is $O^\sim\left(n^3\right)$ to
$O(n^\omega)$ (or $O(n^{1.5}\Omega)$ to $O(\Omega n)$ in the case of
sparse procedures). This
could enable us to recover $O\left(n^{3-\omega}\right)$ ($O(n^{0.5})$)
bits of the determinant by
running modular routines without exceeding the cost of one linear
system solving. This adaptive solution is already implemented in the
dense version of the algorithm, which motivated us to run it on
potentially unsuitable matrices. When comparing the times for the CRA
algorithm and our algorithm applied to sparse matrices (however,
without exploiting their sparsity) we decided that there is a need for
a ''sparse'' version of our algorithm, which will take into account
the sparse structure of the matrices in the subroutines.

In what follows we will shortly present the sparse counterparts of the
subroutines used, give their complexities and discuss some
modification of the parameters if needed. We assume that the cost of
one matrix-vector product is $\Omega=O\left(n\right)$.

Instead of the dense LU, sparse elimination can be used in practice
e.g. for extremely sparse matrices \cite{jgd:2002:villard}.
In general, black box method are preferred. The idea is to precondition
the matrix so that its characteristic polynomial equals its minimal
polynomial \cite{Eberly:1997:RLA,Chen:2002:EP}; and then to compute
the minimal polynomial via Wiedemann's algorithm
\cite{Wiedemann:1986:SSLE}. The complexity of the sparse modular
determinant computation is then $O\left(\Omega n\right)$ \cite[Table
 4]{jgd:2002:villard}. Adaptive solutions exist \cite{duran2003}. 

For solving a sparse system of linear equations the solver of \cite{EbGies2006} can be
used. By similar reasoning as in \cite{Mulders1999}, the cost of solving $Ax=b$ for a sparse matrix $A$ is that of
$O\left(n^{1.5}\log(n\|A\|)+n^{0.5}\log(\|b\|)\right)$ matrix-vector
products and
$O\left(n^{2}\log(n\|A\|)\left(n^{0.5}+\log(\|A\|)\right) + n^2\log(\|b\|)\log(n\|A\|)+ n\log^2(\|b\|)\right)$	
additional arithmetic operations.

If $\|b\|$ is  $O\left(n^{0.5}\right)$ in size and $\Omega=O(n)$, this means
that the complexity of computing $s_n$ is $O\big(n^{2}\log(n\|A\|)(n^{0.5}+\log(\|A\|))\big)$
bit operations.

Currently known sparse determinant algorithms that can be used in the
worst-case step include the CRA loop (with the complexity $O\left(\Omega n
\log(|\det(A)|)\right)$) and the algorithm of \cite{Eberly2000}. By moving to the sparse solver in \cite{Eberly2000} we can
 obtain an algorithm with the worst time complexity of
$O\big( n^{3}\log^{1.5}(n\|A\|)\log\left(\|A\|\right)\log^{2}\left(n\right) \big)$.

All in all, by moving to the sparse procedures, we obtain the
algorithm with the complexity $O\hspace{-1pt}\left(\min\left(k\left(\Omega n^{1.5}\log(n\|A\|)+
n^{2.5}\log(n\|A\|)\log(\|A\|)\right), n^2\log(n\|A\|)\Omega\right)\right)$ where $k$ is the
number of invariant factors. In the propitious case where $k$ is
smaller than $O\left(\sqrt{n}\right)$ 
we obtain an algorithm with the running time better than the currently
known algorithms. 

\subsection{More adaptivity}\label{sec:adaptive}
We start with a simple remark. For every matrix, with each step, the size of $s_{n-i}$
decreases whilst the cost of its computation increases.
In Table \ref{table}, this accounts for better performance of Abbott's algorithm, which
computes only $s_n$, in the case of small $n$. For bigger $n$
calculating $s_{n-1}$ starts to pay out. The same pattern repeats in
further iterations.

The switch between winners in Table \ref{table} can be explained by the fact that,
in some situations, obtaining $s_{n-i}$ by $LU$-factorization
(which costs $\frac{\log \left(s_{n-i}\right)}{\log\left(l\right)}$ the time
of LU) outperforms system solving. Then, this also holds for all
consecutive factors and the algorithm based on CRA wins.
The condition can be checked
{\em a posteriori} by approximating the time of LUs needed to compute the
actual factor. We can therefore construct a condition that would
allow us to turn to the CRA loop in the appropriate moment. This can
be done by changing the
condition in line 27 ($\tilde{\pi}_{i}=\tilde{\pi}_{i-1}$) to
$$
\log\left(\frac{\tilde{\pi}_{i}}{\tilde{\pi}_{i-1}}\right)\leq \frac{time\left(solving\right)}{time\left(LU\right)}\log\left(l\right),
$$
if the primes used in the CRA loop are greater than $l$.
This would result with a performance close to the best and yet flexible.

If, to some extend, $s_{n-i}$ could be approximated {\em a priori},
this condition could be checked before its calculation. This would
require a partial factorization of $s_{n-i+1}$ and probability
considerations as in
section \ref{number}
 and \cite{Eberly2000}.

\section{Conclusions}

In this paper we have presented an algorithm computing the determinant of an
integer matrix. 
In the dense case we proved that the expected complexity of our
algorithm 
is 
$O\big(n^3 \log^2(n\|A\|) \log^{0.5}\left(n\right)\big)$ 
and depends mainly on the cost of the system solving procedure used and the
expected number of invariant factors.
Our algorithm uses an introspective approach so that its
actual expected complexity is only $O\left(n^{3}\left(\log(n) +
\log(\|A\|)\right)^{2}k\right)$ if the number $k$ of invariant
factors is smaller than {\em a priori} expected but greater than
$i_{min}$; 
The actual running time can be even smaller, assuming that any under-estimation resulting from probabilistically correct procedures can be compensated sooner than expected. Moreover, the adaptive approach allows us to switch to the algorithm
with best worst case complexity if it happens that the number of
nontrivial invariant factors is unexpectedly large.
This adaptivity, together with very fast modular routines, allows us to
produce an algorithm, to our knowledge,
faster by at least an order of magnitude than other implementations.

Ways to further improve the running time are to reduce the number of
iterations in the solvings or to group them in order to get some block
iterations as is done e.g. in \cite{Chen:2005:BBC}.
A modification to be tested, is to try to reconstruct $s_n$ with
only some entries of the solution vector
$x=\mathbf{n}/d$.

Parallelization can also be considered to further modify the
algorithm.
Of course, all the LU iterations in one CRA step can be done in
parallel. An equivalently efficient way is to perform several
$p$-adic liftings in parallel, but with less iterations \cite{Dumas2002}.
There the issue is to perform an optimally distributed early
termination.

\renewcommand\refname{References}

\appendix
\begin{appendix}
\section{Properties of matrices with almost uniformly distributed entries}
In this appendix we present some probabilistic properties of matrices
with entries almost uniformly distributed modulo $p^l$, $l\in
\mathbb{Z}$. We consider the case, when the entries are randomly and
uniformly chosen from a set of $S$ contiguous integers $\mathcal{S}=\{a,a+1\dots a+S-1\}$, for any $a$.
As the result, the probability that an entry is equal to a given $d$
modulo $p^l$ is bounded as follows
\begin{equation}
\frac{1}{S}\lfloor \frac{S}{p^l}\rfloor\leq \mathcal{P}(x: x=d \mod p^l) \leq \frac{1}{S}\lceil \frac{S}{p^l}\rceil.
\end{equation}
We set 
\begin{equation}\label{eq:beta}
\beta=\frac{1}{S}\lceil \frac{S}{p^l}\rceil,\quad 
\alpha=\frac{1}{S}\lfloor \frac{S}{p^l}\rfloor.
\end{equation}
This special case of non-uniformly distributed random variables was
widely considered in the thesis of Z. Wan (see \cite{Wan2005}) for $l=1$. In the
following we will first consider the rank modulo $p$ of a matrix under
certain conditions (lemma \ref{app:rankp}).
Then we give the analogues of the theorems 5.9-5.15 of 
\cite{Wan2005} in the case $l>1$ (lemmas \ref{app:non-uni},\ref{app:lxr}, \ref{app:ax}). This allows us to
prove Theorem \ref{expected} on the expected number of invariant factors and
Theorem \ref{lem:pi}, which gives the expected size of over-approximation of
$\mu_i$ in the case of perturbed matrices.

\begin{lemma}\label{app:rankp}
Let $A$ be a $k\times n$, $k \leq n$ integer matrix with entries
chosen uniformly and randomly form $\mathcal{S}$. The probability
that $\rank_p(A)$, the rank modulo $p$ of $A$, is $j$, $0 < j \leq
k$ is less than or equal to 
\begin{align}\label{app:bound}\nonumber
\mathcal{P}\left(\rank_p(A)= j \right)&\leq
\prod_{i=0}^{j-1}(1-\alpha^{(n-i)})\cdot \beta^{(n-j)(k-j)}\cdot
\left(\frac{1}{1-\beta}\right)^{\max(k-j-1,0)}(1+\beta\dots\beta^{k-j})
\\&\leq \beta^{(n-j)(k-j)}\left(\frac{1}{1-\beta}\right)^{k-j},
\end{align}
where $\alpha=\frac{1}{S}\lfloor \frac{S}{p}\rfloor$ and $\beta=\frac{1}{S}\lfloor \frac{S}{p}\rfloor$.
\end{lemma}

\begin{proof}

The proof is inductive on $k-j$ and $j$.
For $j=0$ and $k\ leq n$ the fact that $\rank_p(A)=0$ means
that all the entries of $A$ are zero modulo $p$, that is 
$$
\mathcal{P}(\rank_p(A)=0) \leq \beta^{nk},
$$
the latter being less than $\beta^{nk}\left(\frac{1}{1-\beta}\right)^k$.

Now, denote by $A_i$ the submatrix of $A$ consisting of $i$ first columns.
For $k=j$ we have 
\begin{align*}
&\mathcal{P}(\rank_p(A_k)=k)=\mathcal{P}(\rank_p(A_k)=k~|~\rank_p(A_{k-1}=k-1))\cdot
\mathcal{P}(\rank_p(A_{k-1}=k-1)\\&= \mathcal{P}(\rank_p(A_1)=1)\prod_{i=2}^{k}\mathcal{P}(\rank_p(A_i)=i~|~\rank_p(A_{i-1}=i-1)).
\end{align*}
To compute $\mathcal{P}(\rank_p(A_i)=i~|~\rank_p(A_{i-1})=i-1)$ we
notice the fact that $\rank_p(A_{i-1})=i-1$ means that we can choose an
$(i-1)\times (i-1)$ non-zero minor of $A_{i-1}$. This means that we can leave
the choice of the corresponding $i-1$ entries of the $i$th column free
and only have to ensure that the remaining subvector of size $n-i+1$
is not equal to some given vector. This gives 
$$
\mathcal{P}(\rank_p(A_i)=i~|~\rank_p(A_{i-1}=i-1)\leq (1-\alpha^{n-i+1})
$$
and in consequence
$$
\mathcal{P}(\rank_p(A_k)=k) \leq \prod_{i=1}^{k} (1-\alpha^{n-i+1}).
$$

Now, assume that for all $(j,k)$ such that  $k-j < M$ the bound
\eqref{app:bound} holds. We consider $\mathcal{P}(\rank_p(A_K)=J)$, 
where $K-J=M>0$. We can rewrite:
\begin{align*}
&\mathcal{P}(\rank_p(A_K)=J) = 
\mathcal{P}(\rank_p(A_K)=J~|~\rank_p(A_{K-1})=J)\cdot \mathcal{P}(\rank_p(A_{K-1})=J)
			\\&+
\mathcal{P}(\rank_p(A_K)=J~|~\rank_p(A_{K-1})=J-1)  \cdot \mathcal{P}(\rank_p(A_{K-1})=J-1).
\end{align*}
To estimate $\mathcal{P}(\rank_p(A_K)=J~|~\rank_p(A_{K-1})=J-1)$, as in
previous reasoning, we only have to ensure that $n-J+1$ entries of the
last column are not equal to a certain vector. On the contrary, for
$\mathcal{P}(\rank_p(A_K)=J~|~\rank_p(A_{K-1})=J)$ we notice that we
can leave the choice of $J$ entries corresponding to a non-zero minor
free, but the remaining $n-J$ entries have to be determined modulo
$p$. By induction, we have  
\begin{align*}
&\mathcal{P}(\rank_p(A_K)=J) \leq (1-\alpha^{n-J+1})\cdot \prod_{i=0}^{J-2}(1-\alpha^{(n-i)})\cdot \beta^{(n-J+1)(K-J)}\left(\frac{1}{1-\beta}\right)^{K-J-1}
\\&+ \beta^{n-J}\prod_{i=0}^{J-1}(1-\alpha^{(n-i)})\cdot
\beta^{(n-J)(K-J-1)} \cdot
\left(\frac{1}{1-\beta}\right)^{K-J-1}(1+\beta\dots\beta^{k-j-1}) \\&=
\prod_{i=0}^{J-1}(1-\alpha^{(n-i)})\cdot
\beta^{(n-J)(K-J)}\left(\frac{1}{1-\beta}\right)^{K-J-1}(1+\beta\dots\beta^{K-J})
\end{align*}
which finishes the proof.
\end{proof}

Let us consider the example of $n\times 2$ $\{0,1\}$ matrices. We will
consider $\alpha=\beta=\frac{1}{2}$. We can construct $2^{2n}$
different matrices, $3\cdot (2^{n}-1)$ of which fulfill the
condition that the rank is equal to $1$. The probability of choosing at random a matrix
of rank $1$ is thus $\frac{3(2^n-1)}{2^{2n}}$. The bound given by
Eq. \eqref{app:bound} is $(1-\left(\frac{1}{2}\right)^n)\cdot
\left(\frac{1}{2}\right)^{n-1}\cdot(1+\frac{1}{2})$ which gives
exactly the same value.

The following lemma gives analogues to lemmas 5.10, 5.11 in
\cite{Wan2005} in the case of the ring $\Z_{p^l}$. It proves
that the  vectors of elements from $\mathcal{S}$ can
also be treated as almost-uniformly distributed.

\begin{lemma}\label{app:non-uni}
\begin{enumerate}[(i)]
\item \label{2}Let $t$ be a non-zero mod $p$ vector of size $n$, $d\in
\Z_{p^l}$. Then the probability that a random vector $x\in
\mathcal{S}^n$ is chosen such that $t\cdot
x=d\mod p^l$ is
$$
\mathcal{P}\left(x: t\cdot x=d\mod p^l\right)\leq\frac{1}{S}\lceil\frac{S}{p^l}\rceil
$$ 
\item \label{3}Let $A\in \Z^{m\times n}$, a matrix of rank $r$ such
that the local Smith form of $A$ at $p$ is trivial, $b\in \mathbb{Z}_{p^l}^m$ be given. Then the probability that a random vector $x\in
\mathcal{S}^{m}$ is chosen such that $Ax=b\mod p^l$ is 
$$
\mathcal{P}\left(x: Ax=b\mod p^l\right)\leq \left(
\frac{1}{S}\lceil\frac{S}{p^l}\rceil\right)^r.
$$
\end{enumerate}
\end{lemma}
\begin{proof}
For \eqref{2} the proof of 5.10 from \cite{Wan2005} carry on.
For \eqref{3} we slightly modify the proof of 5.11 from
\cite{Wan2005}. Since $A$ has a trival Smith form modulo $p$ there
exist two matrices $L,R$, $\det(L),\det(R) \neq 0 \mod
p$, such that $A=L\left[\begin{array}{cc}I_r & 0 \\ 0 &
0\end{array}\right]\left[\begin{array}{c}R'\\R''\end{array}\right]$, where $R'=[R_{ij}]$ mod $p$
is a $r\times m$ matrix. We may
therefore transform
$$
\PP(x: Ax=b)=\PP(x: R'x=[L^{-1}b]_{1..r})
$$
Since the determinant of  $R$ is non-zero modulo $p$ there exist a
$r\times r$ minor $R_1$ which is non-zero modulo $p$. This means
that we can find elements $r_{1i_1}\dots r_{1i_r}$ of $R_1$, where $i_k$
are pairwise distinct, such that $r_{ki_k}$ are  non-zero modulo
$p$. Let $d=L^{-1}b$. The probability can be further rewritten:
\begin{align}\label{eq:rx}\nonumber
\PP&(x: R'x=[d]_{1..r})=\sum_{j_1\in\Z_{p^l}}\dots
\sum_{j_r\in\Z_{p^l}}\\
&\PP\hspace{-2pt}\left(\begin{array}{c}
				R_{11}x_1+\dots +\hat{R_{1i_1}x_{i_1}}+
\dots R_{1n}x_n=j_1\\
				R_{21}x_1+\dots +\hat{R_{2i_2}x_{i_2}}+
\dots R_{2n}x_n=j_2\\
				\dots \\
				R_{r1}x_1+\dots +\hat{R_{ri_r}x_{i_r}}+
\dots R_{rn}x_n=j_r\end{array}\right)
			      \PP\hspace{-2pt}\left(\begin{array}{c}
				x_{i_1}=(d_1-j_1)R_{1i_1}^{-1}\\
				x_{i_2}=(d_2-j_2)R_{2i_2}^{-1}\\
				\dots \\
				x_{i_r}=(d_r-j_r)R_{ri_r}^{-1}\end{array}\right)\leq \left(
\frac{1}{S}\lceil\frac{S}{p^l}\rceil\right)^r
\end{align}
We use $\hat{R_{ki_k}x_{i_k}}$ to denote that the element with index
$k$ is omitted in the sum. 
\end{proof}

The following lemmas show that matrices of elements of 
$\mathcal{S}$ can be treated as almost uniformly distributed. 

\begin{lemma}\label{app:lxr}
Let $L,R\in \Z^{n\times n}$ be matrices such that
$|\det\left(L\right)|=|\det\left(R\right)|=1$. Let $\mathcal{I}$,
$\mathcal{J}$ be any disjoint subsets of $\{1\dots n\}^2$ (sets of index pairs). Let $d_{ij}, (i,j)\in \mathcal{I}$ (resp. $D_{st}, (s,t)\in\mathcal{J}$)
be any values (resp. subsets) from $Z_{p^l}$. We consider the probability of choosing a random
matrix $X$ such that $\left(LXR\right)_{ij}=d_{ij}$ for $\left(i,j\right)\in
\mathcal{I}$ under the condition that $(LXR)_{st}\in D_{st}$ for
$\left(s,t\right)\in \mathcal{J}$. We
have
$$
\mathcal{P}\left(\left(LXR\right)_{ij}=d_{ij}~|~\left(LXR\right)_{st}\in D_{st}\right)\leq \left(\frac{1}{S}\lceil\frac{S}{p^l}\rceil\right)^{|\mathcal{I}|}.
$$
\end{lemma}

\begin{proof}

\begin{align}\label{eq:cond}
&\mathcal{P}\left(\left(LXR\right)_{ij}=d_{ij}~|~\left(LXR\right)_{st}\in D_{st}\right)
=\frac{\mathcal{P}\left(\left(LXR\right)_{ij}=d_{ij}\wedge\left(LXR\right)_{st}\in
D_{st}\right)}{\mathcal{P}\left(\left(LXR\right)_{st}\in
D_{st}\right)}
\end{align}
Let $\mathcal{D'}$ denote a set of all possible matrices $[a_{lk}]$ such that $a_{lk}\in D_{lk}$ if
$(l,k)\in\mathcal{J}$ and $a_{lk}\in\mathcal{S}$ otherwise.
Let $\mathcal{D}$ denote a set of matrices from $\mathcal{D'}$ for
which additionally $a_{lk}=d_{lk}$ if $(l,k)\in \mathcal{I}$. Then
Eq. \eqref{eq:cond} can be transformed to
\begin{align*}
\frac{\mathcal{P}\left(X \in
L^{-1}\mathcal{D}R^{-1}\right)}{\mathcal{P}\left(X\in
L^{-1}\mathcal{D'}R^{-1}\right)}
\end{align*}
Notice, that sets $\mathcal{D}$ and $L^{-1}\mathcal{D}R^{-1}$ (resp. $\mathcal{D'}$ and $L^{-1}\mathcal{D'}R^{-1}$) have the
same number of elements. To compute the probability it suffices to
count the number of elements in $\mathcal{D}$ and $\mathcal{D'}$. The
proportion is determined by the choice of elements from $\mathcal{I}$
and is therefore less than or equal to $\left(\frac{1}{S}\lceil\frac{S}{p^l}\rceil\right)^{|\mathcal{I}|}$.

\end{proof}

The methods used to prove Lemmas \ref{app:non-uni} and \ref{app:lxr}
can applied to prove the following lemma.
\begin{lemma}\label{app:ax}
Let $A\in \Z^{m\times n}$ be a matrix such that
the Smith form of $A$ is trivial and $\rank(A)=m\leq n$. Let $\mathcal{I}$,
$\mathcal{S}$ be any disjoint subsets of $\{1\dots m\}^2$ (sets of
index pairs). Let $b_{ij}, (i,j)\in \mathcal{I}$ ($B_{st}, (s,t)\in\mathcal{S}$)
be any values (resp. subsets) from $Z_{p^l}$. We consider the probability of choosing a random
matrix $X$ such that such that $\left(AX\right)_{ij}=b_{ij}$ for $\left(i,j\right)\in
\mathcal{I}$ under the condition that $(LXR)_{st}\in B_{st}$ for
$\left(s,t\right)\in \mathcal{S}$. We
have
\begin{equation}\label{eq:ax}
\mathcal{P}\left(\left(AX\right)_{ij}=b_{ij}~|~\left(AX\right)_{st}\in B_{st}\right)\leq \left(\frac{1}{S}\lceil\frac{S}{p^l}\rceil\right)^{|\mathcal{I}|}.
\end{equation}
\end{lemma}
\begin{proof}
Let matrices $L,R=R'$ be as in the proof of \ref{app:non-uni}. As in
the proof of \ref{app:lxr}, we construct the sets of matrices
$\mathcal{D}$, $\mathcal{D'}$. We have
$$
\mathcal{P}\left(\left(AX\right)_{ij}=b_{ij}~|~\left(AX\right)_{st}\in B_{st}\right)
=\frac{\mathcal{P}\left(RX \in
L^{-1}\mathcal{D}\right)}{\mathcal{P}\left(RX\in
L^{-1}\mathcal{D'}\right)}=\frac{\prod_{i=1\dots m} RX_i\in L^{-1}\mathcal{D}_i}{\prod_{i=1\dots m} RX_i\in L^{-1}\mathcal{D'}_i},
$$
where $X_i$ denote the $i$th column of $X$ and $\mathcal{D}_i
(\mathcal{D'}_i)$, the set of all possible $i$th columns for matrices
from $\mathcal{D} (\mathcal{D'})$. Since \eqref{eq:rx} holds for every vector $L^{-1}d$ of
$L^{-1}\mathcal{D}_i (resp. L^{-1}\mathcal{D'}_i)$, again, we can link the the probability
to the number of elements in $\mathcal{I}$ and conclude that
\eqref{eq:ax} holds.
\end{proof}

We conclude with the following lemma.
\begin{lemma}\label{app:determinant}
Let $V$ be an $k\times n$ matrix, $k\leq n$, such that the Smith form of $V$ is
trival and $V$ has a full rank. Let $M$ be an $n\times k$ matrix
with entries chosen randomly and uniformly from set
$\mathcal{S}$, the probability that $p^{l}< S$ divides the
determinant $\det(VM)$ is at most $\frac{3}{p^{l}}$.
\end{lemma}

\begin{proof}
To check whether $\operatorname{ord}_p\left(\det\left(M\right)\right)\geq l$ we will consider a process of
diagonalization for $M(0)=VM$ mod $p^{l}$ as described in Algorithm {\em
LRE} of \cite{Dumas2001}. It consists of diagonalization and
reduction steps. At the $r$-th diagonalization step, if an invertible entry is found, it is
placed in the $\left(r,r\right)$ pivot position and the $r$th column is
zeroed. If no invertible entry is found, we proceed with a reduction
step i.e. we consider the remaining $\left(n-r+1,n-r+1\right)$ submatrix divided by
$p$. The problem now reduces to determining whether $\operatorname{ord}_p$ of an
$\left(n-r+1,n-r+1\right)$ matrix is greater than $l-n+r-1$.

We can consider matrix $M(0)=M_0+pM_1+p^2M_2\dots +p^{l}M_{l-1}$, where
matrix $M_k\in \Z_p^{n\times n}$, $k=0\dots l-2$ and $M_{l-1}\in
\Z^{n\times n}$. The probability that an entry of $M_k$ is
equal to a certain $d$
modulo $p$ is less that or equal $\frac{1}{N_k}\lceil
\frac{N_k}{p} \rceil$, where $N_k$ is equal to $\lceil
\frac{S}{p^k}\rceil$ by Lemma \ref{app:ax}.

In the process of diagonalization we can find matrices $L_0$, $R_0$,
$\det\left(L_0\right)=\det\left(R_0\right)=1$ such that
$L_0M_0R_0=\diag\hspace{-3pt}\left(\underbrace{1\dots 1}_{r},0 \dots 0\right)$ and
$L_0MR_0=\diag\hspace{-3pt}\left(\underbrace{1\dots 1}_{r},pL_0M_1R_0+\dots\hspace{-2pt} \right)$. Then after the reduction step we
set $M_0\left(1\right)=[(L_0M_1R_0)_{ij}]_{i=r+1\dots n, j=r+1\dots n}$
and $M_k\left(1\right)$ equal to $[(L_0M_{k+1}R_0)_{ij}]_{i=r+1\dots n, j=r+1\dots n}$, $M\left(1\right)=M_0\left(1\right)+pM_1\left(1\right)+\dots$ and we repeat
the diagonalization phase. By construction, the choice of $L_0,
\dots L_{k-1}, R_0,\dots R_{k-1}$ means that certain entries of $M$
are fixed and places us in the situation of Lemmas \ref{app:lxr},\ref{app:ax}. Thanks to that
we can consider the distribution of entries of $M(k)$ as non-uniform
i.e. $\mathcal{P}\left(M(k)_{ij}=d_{ij} \mod p^\alpha~|~L_0\dots
L_{k-1}, R_0\dots R_{k-1}\right)\leq  \frac{1}{N_k}\lceil
\frac{N_k}{p^\alpha} \rceil$. 

Another way to see this is to think of the diagonalization as the
modification to $a_{22}$ in the form of $a_{22}-\frac{a_{21}}{a_{11}}a_{12}$
with $a_{11}$ and $a_{12}$ fixed by the previous step. Then one has
one degree of freedom, say for $a_{21}$ and then $a_{22}$ has to be fixed.

We need only to consider $l-2k$ reductions steps as each reduction is
performed on a matrix of order at least 2 and divides the determinant
by at least $p^2$. Since $k$ is less than  $\lceil l/2 \rceil-1$ and $l\leq \log_p\left(S\right)$, we have $N_k \geq \frac{S}{p^k}
\geq \sqrt{S}$ and since we only consider $p^\alpha < N_k$ we have
$$
\beta_\alpha\left(k\right)=\frac{1}{N_k}\lceil \frac{N_k}{p^\alpha} \rceil\leq
\frac{N_k+p^\alpha-1}{p^\alpha N_k} \leq \frac{2p^\alpha}{p^\alpha\left(p^\alpha+1\right)}=\frac{2}{p^\alpha+1}$$
throughout the process. We therefore
now set $\beta_\alpha=\frac{2}{p^\alpha+1}$ and use it as a bound for
$\beta_\alpha\left(k\right)$, $k=1,2\dots$ in our calculations. 

The proof is inductive on $n$, the dimension of the matrix $M\left(k\right)$ and $l$, the
current exponent. We fix the diagonalization/reduction matrices
$L_0\dots L_{k-1},R_0\dots L_{k-1}$ and consider the conditional
probability $\mathcal{P}^{k-1}=\mathcal{P}\left(\cdot ~|~L_0\dots L_{k-1},R_0\dots R_{k-1}\right)$.

First, for $l=1$, \cite[Thm 5.13]{Wan2005} gives
$$
\mathcal{P}^{k-1}\left(p\nmid \det\left(M(k)\right)\right)\leq \prod_{i=1}^{n}\left(1-\beta^i_1\right).
$$

This transforms to
\begin{equation}\label{eq:det}
\mathcal{P}^{k-1}\left(p\mid \det\left(M\left(k\right)\right)\right)\leq \sum_{i=1}^{n}\beta^i_1\leq
\frac{\beta_1}{1-\beta_1}.
\end{equation}
Thus, the probability can be bounded by $\min\left(1,\frac{2}{p-1}\right)$ and
therefore by $\frac{3}{p}$.

For $n>1$ we will sum over all possible choices of $L_k$ and $R_k$. We
will divide the sum on the cases when applying $L_k$ and $R_k$ leads
to the diagonalization of at least $r$ entries. We call such an event
$E_r$.

Then for $n=2, l=2$:
\begin{align*}
&\mathcal{P}^{k-1}\left(p^2\mid \det\left(M\left(k\right)\right)\right)\leq
\sum_{L_k,R_k \in E_1} \mathcal{P}^{k-1}\left(p^2 |
\left(L_kM\left(k\right)R_k\right)_{22}~|~L_k,R_k\right)\mathcal{P}^{k-1}\left(L_k,R_k\right)
\\&+\mathcal{P}^{k-1}\left(p | M\left(k\right)_{ij}, i,j=1,2\right)\leq  \beta_2 + \beta_1^{4}\leq \frac{2}{p^2+1}+\left(\frac{2}{p+1}\right)^4 \leq \frac{3}{p^2}
\end{align*}

Now we suppose inductively that $\mathcal{P}^{k-1}\left(p^i\mid \det\left(M\left(k\right)\right)\right)\leq \frac{3}{p^i}$
for all $i<l$.
Then for $n=2, 2 < l < n$ the induction gives 
\begin{align*}
&\mathcal{P}^{k-1}\left(p^l\mid \det\left(M\left(k\right)\right)\right)\leq
\sum_{L_k,R_k \in E_1} \mathcal{P}^{k-1}\left(p^l |
\left(L_kM\left(k\right)R_k\right)_{22}~|~L_k, R_k\right)\mathcal{P}^{k-1}\left(L_k,R_k\right)\\
&+\mathcal{P}^{k-1}\left(p | M\left(k\right)_{ij},
i,j=1,2\right)\mathcal{P}^{k-1}\left(p^{l-2}|\det\left(M\left(k+1\right)\right)\right)
\leq \beta_l\left(k\right) + \beta_1\left(k\right)^4\frac{3}{p^{l-2}}.
\end{align*}
%
The latter is less than
$\frac{3}{p^l}$ when
\begin{equation}\label{eq:P2l}
\beta_1\left(k\right)^4 3p^2 \leq 1.
\end{equation}
With $\beta_1\left(k\right)\leq \frac{2}{p+1}$ this means that
$\frac{48}{\left(1+1/p\right)^2\left(p+1\right)^2}=\frac{48}{\left(p+2+1/p\right)^2}\leq 1$ which is fulfilled for $p>3$.
For primes $p=2,3$ we have to use a sharper bound for
$\beta_l\left(k\right)$. Since $p^l < N_k$ and $l > 2$ we have 
\begin{equation}\label{eq:sharpbd}
\beta_1\left(k\right) \leq \frac{p^l+1+p-1}{\left(p^l+1\right)p} \leq
\frac{p^{l-1}+1}{p^l+1} < \frac{p+1}{p^2+1}.
\end{equation}
This allows us to prove the inequality
\eqref{eq:P2l} for $p=3$ since $\left(\frac{2}{5}\right)^4 27 < 0.7$. For $p=2$
and $l>3$ also $\left(\frac{9}{17}\right)^4 12 < 0.95$ . For the remaining
case $p=2$, $l=3$ we can bound $\mathcal{P}^{k-1}\left(p\mid \det\left(M\left(k+1\right)\right)\right)$ by $1$ instead of
$\frac{3}{2}$ and then one can prove that
$\beta_3\left(k\right)+\beta_1\left(k\right)^4\leq \frac{2}{9} + \left(\frac{5}{9}\right)^4 < 0.32 \leq\frac{3}{8}$.


Now we will consider $n > 2$. Again we can sum
over all possible diagonalization and reduction steps combinations and the
resulting bound for the probability is
\begin{align}\label{eq:lln}\nonumber
&\mathcal{P}^{k-1}\left(p^l\mid \det\left(M\left(k\right)\right) \right)\leq
 \mathcal{P}^{k-1}\left(p |
M\left(k\right)_{ij}\forall_{i,j \leq n}\right)\\\nonumber&+
\sum_{r=1}^{n-l}\sum_{L_k,R_k\in E_r} \mathcal{P}^{k-1}\left(p |
\left(L_kM\left(k\right)R_k\right)_{ij}\forall_{i,j \leq n-r}~|~L_k,R_k\right)\mathcal{P}^{k-1}\left(L_k,R_k\right)\\\nonumber&+
\sum_{r=n-l+1}^{n-2}\sum_{L_k,R_k\in E_r} \mathcal{P}^{k-1}\left(p |
\left(L_kM\left(k\right)R_k\right)_{ij}\forall_{i,j \leq
n-r}~|~L_k,R_k\right)\cdot \\\nonumber
&\hspace{110pt}\mathcal{P}^{k-1}\left(L_k,R_k\right)\mathcal{P}^{k}\left(p^{l-n+r} | \det\left(M(k+1)\right)\right)\\\nonumber&+
\sum_{L_k,R_k\in E_{n-1}}\mathcal{P}^{k-1}\left(p^l | \left(L_kM\left(k\right)R_k\right)_{nn}~|~L_k,R_k\right)\mathcal{P}^{k-1}\left(L_k,R_k\right)
\leq \sum_{i=l}^{n}\beta_1\left(k\right)^{i^2}\\&+\sum_{r=n-l+1}^{n-2}\sum_{L_k,R_k\in E_r} \beta_1\left(k\right)^{\left(n-r\right)^2}\mathcal{P}^{k-1}\left(L_k,R_k\right)\mathcal{P}^{k}\left(p^{l-n+r} | \det\left(M\left(k+1\right)\right)\right)+\beta_l\left(k\right)
\end{align}
for $l \leq n$ and similarly for $l > n$
\begin{align}\label{eq:lgn}\nonumber
&\mathcal{P}^{k-1}\left(p^l\mid \det\left(M\left(k\right)\right)\right)\leq
 \mathcal{P}^{k-1}\left(p | M\left(k\right)_{ij}\forall_{i,j \leq n}\right)\mathcal{P}^{k-1}\left(p^{l-n}| \det\left(M(k+1)\right)\right) \\\nonumber&+
\sum_{r=1}^{n-2}\sum_{L_k,R_k\in E_r} \mathcal{P}^{k-1}\left(p |
\left(L_kM\left(k\right)R_k\right)_{ij}\forall_{i,j \leq n-r}~|~L_k,R_k\right)\cdot \\\nonumber
&\hspace{95pt}\mathcal{P}^{k-1}\left(L_k,R_k\right)\mathcal{P}^{k}\left(p^{l-n+r} | \det\left(M(k+1)\right)\right)\\\nonumber&+
\sum_{L_k,R_k\in E_{n-1}}\mathcal{P}^{k-1}\left(p^l | \left(L_kM\left(k\right)R_k\right)_{nn}~|~L_k,R_k\right)\mathcal{P}^{k-1}\left(L_k,R_k\right)\\\nonumber&\leq \beta_1\left(k\right)^{n^2}\mathcal{P}^{k-1}\left(p^{l-n}| \det\left(M\left(k+1\right)\right)\right)+\beta_l\left(k\right)\\&+ \sum_{r=1}^{n-2}\sum_{L_k,R_k\in E_r} \beta_1\left(k\right)^{\left(n-r\right)^2}\mathcal{P}^{k-1}\left(L_k,R_k\right)\mathcal{P}^{k-1}\left(p^{l-n+r} | \det\left(M\left(k+1\right)\right)~|~L_k,R_k\right).
\end{align}

Again, we can use the induction to get the bound $\mathcal{P}^{k}\left(p^{l-i}\mid
\det\left(M(k+1)\right)\right)\leq \frac{3}{p^{l-i}}$. Then, we can then bound
both sums by
\begin{equation}\label{eq:Pln}
\mathcal{P}^{k-1}\left(p^l\mid \det\left(M\left(k\right)\right) \right) \leq
\sum_{i=2}^{\infty}\beta_1\left(k\right)^{i^2}\frac{3}{p^{l-i}} + \beta_l\left(k\right)
\leq \frac{3\beta_1\left(k\right)^4}{p^{l-2}}\frac{1}{\left(1-\beta_1\left(k\right)^5p\right)} + \beta_l\left(k\right).
\end{equation}
To prove the inequality $\mathcal{P}^{k-1}\left(p^l\mid \det\left(M\left(k\right)\right) \right) \leq \frac{3}{p^l}$, we have to consider several cases. For $p > 3$ we use the bound $\beta_1$ and $\beta_l$ for
$\beta_1\left(k\right)$ and $\beta_l\left(k\right)$ respectively. Then we have
\begin{align*}
&\frac{3\cdot 2^4p^2\left(p+1\right)}{p^{l}\left(\left(p+1\right)^5-p2^5\right)} +
\frac{2}{p^l+1} < \frac{2}{p^l}+
\frac{1}{p^l}\frac{48p^2\left(p+1\right)}{\left(p+1\right)^5 - \left(p+1\right)2^4} <
\frac{2}{p^l}+\frac{1}{p^l}\frac{48p^2}{\left(p+1\right)^4 - 2^4}\\ & <
\frac{2}{p^l}+\frac{1}{p^l}\frac{48p^2}{25p^2+4\cdot 5p^2+6\cdot p^2+4 \cdot5+1-16}
< \frac{2}{p^l}+\frac{1}{p^l}\frac{48p^2}{51p^2} < \frac{3}{p^l}.
\end{align*}
For $p=3$ it can be explicitly checked that $\mathcal{P}^{k-1}\left(p^l\mid \det\left(M\right)\right) <
\frac{3}{p^l}$ using the bound $\frac{p+1}{p^2+1}$ for $\beta_1\left(k\right)$ (notice that
$N^k > p^l $). In this case we get $\frac{1}{3^l}\frac{3\left(\frac{2}{5}\right)^4
3^2}{\left(1-\left(3\frac{2}{5}\right)^5\right)} + \frac{2}{3^l} < \frac{1}{3^l}2.75$.

For $p=2$ we have to consider $2^2, 2^3, 2^4$ and $2^l$ for $l>4$
separately and use the sharper bound from Eq. \eqref{eq:sharpbd}. Let us rewrite \eqref{eq:lln} and \eqref{eq:lgn}
in this cases.
\begin{itemize}
\item $l=2$:
$$\mathcal{P}^{k-1}\left(2^2\mid \det\left(M\left(k\right)\right)\right)\leq
\sum_{i=2}^{n}\beta_1\left(k\right)^{i^2} + \beta_2\left(k\right) \leq
\beta_1\left(k\right)^4\frac{1}{1-\beta_1\left(k\right)^5} + \beta_2\left(k\right).$$
As $\beta_1\left(k\right) \leq \frac{2+1}{4+1}$ we have $0.65 < 0.75$.
\item $l=3$:
\begin{align*}
\mathcal{P}^{k-1}\left(2^3\mid \det\left(M\left(k\right)\right)\right)&\leq
\sum_{i=3}^{n}\beta_1\left(k\right)^{i^2}+\beta_1\left(k\right)^{4}\cdot
1
+ \beta_3\left(k\right)
\\&\leq \beta_1\left(k\right)^9\frac{1}{1-\beta_1\left(k\right)^7}
+\beta_1(k)^{4}
+ \beta_3(k).
\end{align*}
As $\beta_1\left(k\right) \leq \frac{4+1}{8+1}$ we have $0.33 < 0.375$.
\item $l=4$:
\begin{align*}
&\mathcal{P}^{k-1}\hspace{-1.4pt}\left(2^4\mid \det\left(M(k)\right)\right)\hspace{-1pt}\leq\hspace{-1pt} \sum_{i=4}^{n}\beta_1(k)^{i^2}\hspace{-4pt} +
\beta_1(k)^{9}\hspace{-2pt}\cdot 1
+ \beta_1(k)^{4}\mathcal{P}^{k}\left(2^{2}\mid
\det\left(M(k+1)\right)\right)+\beta_4(k)
\\&\leq \beta_1(k)^{16}\frac{1}{1-\beta_1(k)^9}+\beta_1(k)^{9} + \beta_1(k)^{4}\frac{3}{4}+\beta_4(k).
\end{align*}
As $\beta_1\left(k\right) \leq \frac{8+1}{16+1}$ we have $0.18 < 0.1875$.
\item $l>4$:

We use inequality \eqref{eq:Pln} with
$\beta_1\hspace{-1pt}\left(k\right)$ bounded by $\frac{p^4+1}{p^5+1}$.
We get $\mathcal{P}^{k-1}\hspace{-1pt}\left(2^l\mid
\det\left(M\left(k\right)\right)\right)$ is less than $\frac{1}{2^l}\left(\frac{3\left(2^4+1\right)^4
2^2 }{\left(2^5+1\right)^4\left(\left(2^5+1\right)^5\right)-2\left(2^4+1\right)}+2\right) < 2.92\frac{1}{2^l} <  \frac{3}{2^l}$.
\end{itemize}
We have thus proven that $\mathcal{P}^{k-1}\left(p^{l} |
\det(M(k))\right)\leq \frac{3}{p^{l}}$ for every $l >0$ and
every size $n$ of $M(k)$. Thus, $\mathcal{P}(p^l | \det(VM))$ is also
less than or equal $\frac{3}{p^l}$.
\end{proof}

\end{appendix}

\end{document}